\documentclass[twocolumn]{aa}
\usepackage{graphicx}
\usepackage[round]{natbib}
\usepackage{epsfig}
\usepackage{aalongtable}
\begin{document}

\title{Integrated spectrum of the planetary nebula NGC\,7027\thanks{
Tables~2 and 4 are only available in electronic form at 
the CDS via anonymous ftp to cdsarc.u-strasbg.fr (130.79.126.5)
or via http://cdsweb.u-strasbg.fr/cgi-bin/qcat?J/A+A/; 
Tables~1, 3, and Fig.~1 and the
Appendices A and B with their tables are only available
in electronic form at http://www.edpsciences.org}}

\author{Y. Zhang\inst{1}\fnmsep\thanks{Also Space Telescope Science Institute, 3700 San Martin Drive, Baltimore, MD
21218}, X.-W. Liu\inst{1}, S.-G.Luo\inst{1},
D. P{\' e}quignot\inst{2}
\and
M. J. Barlow\inst{3}}

\offprints{Y. Zhang}
\institute{Department of Astronomy, Peking University, Beijing 100871,
     P. R. China\\
     \email{zhangy@stsci.edu}
  \and
      LUTH, Laboratoire l'Univers et ses Th\'eories,
             associ\'e au CNRS (FRE 2462) et \`a l'Universit\'e Paris 7,
             Observatoire de Paris-Meudon, F-92195
             Meudon C\'edex, France
   \and 
    Department of Physics and Astronomy, University College London, Gower Street, London WC1E 6BT, UK\\}

   \date{Received ; accepted }

   \abstract{
We present deep optical spectra of the archetypal young planetary nebula (PN)
\object{NGC\,7027}, covering a wavelength range from 3310 to 9160\,{\AA}.  The
observations were carried out by uniformly scanning a long slit across the
entire nebular surface, thus yielding average optical spectra for the whole
nebula. A total of 937 emission features are detected.
The extensive line list presented here should prove valuable for future
spectroscopic analyses of emission line nebulae. The optical data, together
with the archival {\it IUE}\ and {\it ISO}\ spectra, are used to probe the
temperature and density structures and to determine the elemental abundances
from lines produced by different excitation mechanisms.  Electron temperatures
have been derived from the hydrogen recombination Balmer jump (BJ), 
from ratios of \ion{He}{i} optical recombination lines (ORLs) and from a
variety of diagnostic ratios of collisionally excited lines (CELs). Electron
densities have been determined from the intensities of high-order
\ion{H}{i} Balmer lines and of \ion{He}{ii} Pfund lines, as well as from a 
host of
CEL diagnostic ratios. CEL and ORL diagnostics are found to yield compatible
results.  Adopting respectively electron temperatures of 
$T_{\rm e} = 12\,600$ and 15\,500\,K for ions with ionization potentials 
lower or 
higher than 50\,eV and a constant density of  $N_{\rm e}=47\,000$\,cm$^{-3}$,
elemental abundances have been
determined from a large number of CELs and ORLs.  The C$^{2+}$/H$^+$,
N$^{2+}$/H$^+$, O$^{2+}$/H$^+$ and Ne$^{2+}$/H$^+$ ionic abundance ratios
derived from ORLs are found to be only slightly higher than the corresponding
CEL values.  We conclude that whatever mechanism is causing the BJ/CEL
temperature discrepanies and the ORL/CEL abundance discrepancies that have 
been
observed in many PNe, it has an insignificant effect on this bright young
compact PN.  The properties of the central star are also discussed. Based on
the integrated spectrum and using the energy-balance method, we have 
derived an
effective temperature of 219\,000\,K for the ionizing star.  Finally, we report
the first detection in the spectrum of this bright young PN of 
Raman-scattered \ion{O}{vi} features at
6830 and 7088\,{\AA}, pointing to the existence of abundant neutral hydrogen
around the ionized regions.

   }

\authorrunning{Y. Zhang et al.}
\titlerunning{The planetary nebula NGC\,7027}

   \maketitle
%

\section{Introduction}

\object{NGC\,7027 (PNG~84.9$-$3.4\degr\,1)} is a high-excitation, young, dense
planetary nebula (PN) ionized by a hot central star (CS). Due to its proximity
\citep[$\sim880\pm150$\,pc;][]{masson}, its high surface brightness and 
exceedingly
rich spectrum, \object{NGC\,7027} has been one of the most 
intensively observed
astronomical objects. It has long served as an ideal laboratory for studies
of the physics of low density astrophysical plasmas and for testing the
accuracy of current atomic data. Extensive spectroscopy has been carried
out for this object.  Early spectroscopic observations using photographic
plates were presented by \citet{bowen} and \citet{wyse}. Combining
photographic and spectrophotometric techniques, \citet{kaler} measured
intensities of lines between 3132--8665\,{\AA}. They suggested that the
nebula may contain a dense central ionized region with electron density
of $N_{\rm e}\sim10^7$\,cm$^{-3}$. Using an echelle spectrograph equipped with
 a CCD detector,
\citet{keyes} presented the spectrum of \object{NGC\,7027} over the 
wavelength range
3673--8750\,{\AA} and showed that the nebula has an average electron 
density
$\sim60\,000$\,cm$^{-3}$ and an electron temperature $\sim14\,000$\,K.  
Based on deep
medium resolution optical spectra, \citet{pequignotba1994} detected 
forbidden
lines from heavy elements with $Z>30$ from this nebula, while deep CCD
spectra encompassing the wavelength range of 6540--10460\,{\AA} have been
published by \citet{baluteau}. 

The infrared (IR) spectrum of \object{NGC\,7027} has also been 
extensively studied
\citep[e.g.][]{dyck,telesco,melnick,condal,gee,nagata,rudy}. 
Owing to its status as a wavelength standard, IR observations were 
repeatedly taken by the Infrared Space
Observatory ({\it ISO}) \citep[see e.g.][]{liuiso,salas}. One advantage of
observing in the IR is the reduced effect of dust extinction. The IR spectrum
also gives access to lines emitted by ionized species unobservable in 
other
wavelength regions, thus reducing uncertainties in abundance determinations.
The ISO observations yielded a large number of molecular features, which
give information of the photodissociation regions and molecular envelope
surrounding the ionized region \citep[e.g.][]{liuiso,justtanont}. A large
number of ultraviolet (UV) spectra of \object{NGC\,7027} are available from the
International Ultraviolet Explorer
({\it IUE}) data archive \citep{bohlin,keyes}. The UV spectra are particularly
useful for the investigation of the interstellar and circumstellar reddening,
mass-loss rate and properties of the stellar winds. The UV spectra yield
intensities of strong collisionally excited lines (CELs) from ionized carbon
not available in other wavelength regions, such as \ion{C}{iii}]
$\lambda\lambda$1906,1909 and \ion{C}{iv} $\lambda\lambda$1548,1550, and are
therefore crucial for the determination of carbon abundances.

NGC\,7027 is known to contain a large amount of local dust. \citet{seaton}
shows that the dust reddening varies across the nebula. He used a
two-dust-component model to fit the radio and infrared observations. A
wedge-shaped extinction model is presented by \citet{middlemass}, assuming that
dust extinction increases linearly across the surface of the nebula.
\citet{walton} deduced an extinction map of NGC\,7027 and found an average
extinction of $E_{\rm B-V}=1.02$. They also deduced that the CS lies at a
position of low extinction, $E_{\rm B-V}=0.8$. The latter result is however not
supported by the high-resolution imaging observations of \citet{robberto} and
\citet{wolff} who find that $E_{\rm B-V}=1.07$ and 1.10 towards the CS,
respectively.

The large number of CELs and optical recombination lines (ORLs) detected in the
spectrum of \object{NGC\,7027} from a variety of ionic species provide a unique
opportunity to study the chemical composition of the nebula at a level normally
unachievable in other emission line nebulae.  A long-standing problem in
nebular abundance studies has been that heavy-element abundances derived
from ORLs are systematically higher than those derived from CELs. In extreme
cases, the discrepancies can exceed a factor of 10 (see Liu 2003 for a recent
review). Related to the CEL versus ORL abundance determination
dichotomy, electron temperatures determined from traditional CEL diagnostic
ratios, such as the [\ion{O}{iii}] $(\lambda4959 
+\lambda5007)/\lambda4363$ nebular
to auroral line ratio, are found to be systematically higher than those
determined from analyses of the recombination spectrum, such as the 
\ion{H}{i}
recombination continuum Balmer discontinuity and \ion{He}{i} recombination
lines.  The two phenomena are found to be correlated, the larger the
discrepancy between the CEL and ORL abundances, the larger the difference
between the CEL and recombination line/continuum temperatures
\citep{liuluo,liu}. \citet{liubarlow} ascribed the CEL versus ORL
temperature/abundance determination discrepancies to the existence of ultra
cold H-deficient inclusions embedded in nebulae.  In the case of NGC\,7027,
\citet{salas} compared C$^{++}$/H$^+$, N$^{++}$/H$^+$ and O$^{++}$/H$^+$ 
ionic
abundance ratios derived from CELs and from ORLs and found that the 
abundance
discrepancies in this young, compact PN are insignificant.

Compared to the gaseous nebula, the CS of \object{NGC\,7027} is less well
studied. Embedded in a bright nebula, it is a difficult task to measure its
visual magnitude, which is necessary for the determination of the effective
temperature of the CS using the Zanstra method.  Analyses published so far have
yielded discrepant results. \citet{shaw} estimate that the CS has an effective
temperature of 180\,000\,K, compared to a much higher value of 310\,000\,K
obtained by \citet{walton}.  More recently, based on near IR observations where
the CS is better observed, \citet{latter} derive a value of 198\,000\,K.  All
these estimates were based on the Zanstra method \citep{zanstra}.
\citet{latter} also estimate that the CS has a mass of 0.7 solar masses.  By
comparing the observed nebular elemental abundances and the predictions of
semi-analytical TP-AGB evolutionary models of \citet{marigo}, \citet{salas}
suggest that the CS is probably descended from a C-rich progenitor star with a
main sequence mass between 3--4 solar masses.

In this paper, we present new optical spectra of this archetypal young PN,
obtained by uniformly scanning a long slit across the entire nebular surface.
The spectra reveal a large number of emission lines. The comprehensive line
list generated from the spectra is intended to facilitate future spectroscopic
observations and line identifications of emission line nebulae.  In order to
have a comprehensive view of the thermal and density structures of the nebula
and its chemical composition, we have also included in our analysis available
{\it IUE}\ spectra in the UV and {\it ISO} spectra from the near- to 
far-IR.  Section~2
describes our new optical observations and the procedures of data reduction and
presents identifications and fluxes of detected emission lines.  In Section~3,
we report the first detection of Raman-scattered features from this nebula.
Dust extinction towards \object{NGC\,7027} is briefly discussed in Section~4.
In Section~5, we present plasma diagnostic results.  Ionic and elemental
abundances derived from ORLs and CELs are presented and compared in Section~6.
Section~7 discusses the CS.  A summary then follows in Section~8.

\section{Observations}

\subsection{Optical spectroscopy}

The observations were carried out in 1996 and 1997, using the ISIS long-slit
double spectrograph mounted on the 4.2\,m William Herschel Telescope (WHT) at
La Palma. An observational journal listing the spectral wavelength coverage and
FWHM resolution is presented in Table~\ref{jan}. For both the Blue and Red
arms of the spectrograph, a Tek $1024\times1024$ 24$\mu$m$\times$24$\mu$m chip
was used, yielding a spatial sampling of 0.3576\,arcsec\,per pixel projected on
the sky.  In 1996, a 600\,g\,mm$^{-1}$ and a 316\,g\,mm$^{-1}$ grating were
used for the Blue and Red Arm, respectively. In 1997, they were replaced
respectively by a 1200\,g\,mm$^{-1}$ and a 600\,g\,mm$^{-1}$ grating.  A
dichroic with a cross-over wavelength near 5200\,{\AA} was used to split the
light beam.  In order to avoid uncertainties caused by ionization
stratification when comparing the ionic abundances derived from the optical
spectrum with those deduced from UV and IR spectra obtained with
space-borne facilities which use large apertures and thus yield total line
fluxes for the whole nebula, the long slit of the ISIS spectrograph was used to
uniformly scan across the entire nebular surface by differentially driving the
telescope in Right Ascension.  The mean optical spectrum thus obtained, when
combined with the total H$\beta$ flux measured with a large aperture, $\log
F(\mathrm{H}\beta) = -10.12$ (erg\,cm$^{-2}$\,s$^{-1}$) \citep[][]{shaw},
yields integrated fluxes for the whole nebula for all emission lines detected,
which are therefore directly comparable to those measured with the space-borne
facilities.

Four wavelength regions were observed in 1996. The $\lambda\lambda$3620--4400
and $\lambda\lambda$4200--4980 ranges were observed with the Blue Arm and the
$\lambda\lambda$5200--6665 and $\lambda\lambda$6460--7930 regions with the Red
Arm. In 1997, ten spectral regions covering the wavelength range from 
3310 to
9160\,{\AA} were observed (c.f. Table~\ref{jan}). Spectral lines falling in the
overlapping wavelength region of two adjacent wavelength setups were used 
to
scale the spectra and ensured that all spectra were on the same flux scale.
Short exposures were taken in order to obtain intensities of the brightest
emission lines, which were saturated on spectra with long exposure times.

All spectra were reduced using the {\sc long92} package in {\sc
midas}\footnote{{\sc midas} is developed and distributed by the European
Southern Observatory.} following the standard procedure. Spectra were
bias-subtracted, flat-fielded and cosmic-rays removed, and then wavelength
calibrated using exposures of copper-neon and copper-argon calibration lamps.
Absolute flux calibration was obtained by observing the {\it HST}\
spectrophotometric standard stars, \object{BD\,+28\degr\,4211} and
\object{Hz\,44} using a 6\,arcsec wide slit. 

All line fluxes, except those of the strongest, were measured using Gaussian
line profile fitting. For the strongest and isolated lines, fluxes obtained by
direct integration over the observed line profile were adopted. A total of 937
distinct emission features were measured, including 739 isolated lines, 198
blended features of two or more lines and 18 unidentified features.  Combining
the single and blended features, a total of 1174 lines were identified.
The main references used for line identifications and laboratory wavelengths are
\citet{keyes}, \citet{pequignotba1994}, \citet{baluteau}, 
Hirata $\&$ Horaguchi (1995)\footnote{http://amods.kaeri.re.kr/spect/SPECT.html.},
the NIST Spectroscopic Database\footnote{http://www.physics.nist.gov/cgi-bin/AtData/main\_asd.}, the Atomic Line List
Version 2.04 compiled by P. A. M. van Hoof\footnote{http://www.pa.uky.edu/$\sim$peter/atomic/.} and reference therein.  The
complete optical spectrum is plotted in Fig.~\ref{all} with identified lines
marked.  Several emission lines from ions of $Z>30$ have been detected. In
Appendix~A, their observed fluxes are compared with those obtained by
\citet{pequignotba1994}.

A full list of lines detected in our deep integrated optical spectrum and their
measured fluxes are presented in  Table~2. The first column gives the
observed wavelengths after correcting for Doppler shifts determined from 
H~{\sc
i} Balmer lines. The observed fluxes are given in column 2.  Column 3 lists the
fluxes after correcting for dust extinction (cf.  Section~\ref{redden}).  
The
remaining columns of the table give, in sequence, ionic identification,
laboratory wavelength, multiplet number (with a prefix `V' for permitted lines,
`F' for forbidden lines,  and `H' and `P' for hydrogen Balmer and Paschen
lines, respectively), lower and upper spectral terms of the transition,
and statistical weights of the lower and upper levels, respectively.  All
fluxes are normalized such that $F({\rm H}\beta)= I({\rm H}\beta) = 100$. 

\subsection{{\it IUE}\ and {\it ISO}\ observations} 

\object{NGC\,7027} was observed by the {\it IUE}\ over the period from 1979 to
1983. Both high- and low-resolution spectra were obtained with the Short
Wavelength Prime (SWP) and with the Long Wavelength Redundant (LWR) cameras,
covering the wavelength ranges  1150--1975\,{\AA} and
1910--3300\,{\AA}, respectively. Only observations obtained with the
{\it IUE}\ large aperture are included in the current analysis. They are listed
in Table~\ref{iuejan}. The {\it IUE}\ large aperture has an oval shape of
dimensions $10^{''}\times23^{''}$, larger than the angular size of the ionized
region of \object{NGC\,7027} as revealed by radio imaging
\citep[c.f.][]{roelfsema}.  All the {\it IUE}\ spectra were retrieved from the
{\it IUE}\ Final Archive hosted by the ESA Data Centre in Vilspa, Spain. Short
exposures were used for measuring the strongest lines, which were saturated in
the spectra of long exposure time. The measured line fluxes are given in
Table~4. Our measurements are in reasonable agreement with those
reported by \citet{keyes}, who used fewer spectra (SWP17240L, SWP17242L,
SWP19579L, SWP19877L, LWR05615L, LWR15105L, LWR15861L and LWR15862L).  The
fluxes were normalized to $F({\rm H}\beta)=100$ using an H$\beta$ flux of $\log
F({\rm H}\beta)=-10.12$ (ergs\,cm$^{-2}$\,s$^{-1}$) \citep{shaw} and then
dereddened using a reddening constant of $c({\rm H}\beta)=1.37$ (c.f.
Section~\ref{redden}).  The normalized observed and dereddened line fluxes
are listed in the last column of Table~4.

\object{NGC\,7027} has also been extensively observed with the Short Wavelength
Spectrometer \citep[SWS;][]{graauw} and Long Wavelength Spectrometer
\citep[LWS;][]{clegg}, on board {\it ISO} \citep{kessler}, covering wavelength
ranges from 2.38--45.2\,$\mu$m and 40--197\,$\mu$m, respectively.  The smallest
ISO-SWS aperture has a rectangular size of $14^{''}\times20^{''}$, comparable
to the large aperture of {\it IUE}.  The LWS has an effective entrance aperture
of approximately 70\arcsec\, in diameter.  Thus both the SWS and LWS apertures
are big enough to contain the whole ionized region of NGC\,7027. The LWS and
SWS spectra of NGC\,7027 have previously been analyzed by \cite{liuiso} and
\cite{salas}. Emission line fluxes reported in these papers are included in the
current analysis. We normalized the observed fluxes to $F({\rm H}\beta)=100$
using the total H$\beta$ flux given above and then dereddened assuming $c({\rm
H}\beta)=1.37$.

\section{Raman scattering features} \label{ra}

A broad feature at 4852\,{\AA} detected in the optical spectrum of
\object{NGC\,7027} has been proposed to be the Raman-scattered line of
\ion{He}{ii}(2-8) by atomic hydrogen \citep{pequignotbm}. This was the first
detection of a Raman line from a PN.  More recently, \citet{pequignot2003} 
report
the detection in this PN of another two Raman features, \ion{He}{ii}(2-10) 
and
\ion{He}{ii}(2-6) at 4332\,{\AA} and 6546\,{\AA}, respectively.  Except 
for the
6546\,{\AA} feature, which is seriously blended with the strong [\ion{N}{ii}]
line at 6548\,{\AA}, the detection and identification of the
other two \ion{He}{ii} Raman
lines are confirmed by our new observations (c.f. Fig.~\ref{all}).  The
detection of Raman lines suggests the abundant presence of neutral hydrogen 
around the ionized regions of this high excitation PN.

Our spectrum also reveals an abnormally broad feature redward of the
[\ion{Kr}{iii}] $\lambda6827$ line (Fig.~\ref{raman}). A two-Gaussian profile fit
shows that the broad feature has a central wavelength of 6829.16\,{\AA} and a
$FWHM$ of 253\,km\,s$^{-1}$, which is much broader than the [\ion{Kr}{iii}]
$\lambda6827$ line ($FWHM\sim88$\,km\,s$^{-1}$). The feature was also detected
by \citet{pequignotba1994} and was identified by them as a \ion{Si}{ii} line.
Given the narrow width of other \ion{Si}{ii} lines, we regard this as a
mis-identification. \citet{schmid} suggested that Raman scattering of the
\ion{O}{vi} $\lambda1032,1038$ resonance doublet by neutral hydrogen gives rise
to two velocity-broadened lines at 6830 and 7088\,{\AA} that have been widely
observed in the spectra of symbiotic stars. Therefore, based on its measured
wavelength and the large $FWHM$, we identify the broad emission feature at
6829.16\,{\AA} as the \ion{O}{vi} Raman line at 6830\,{\AA}. 
Our identification is strengthened by the detection of another feature,
marginally above the detection limit, at 7088\,{\AA}, i.e. at the 
expected position of the other O~{\sc vi} Raman line. The measurements 
yield a $\lambda7088/\lambda6830$ intensity ratio of approximately 1/7, smaller 
than the ratio of 1/4 typically found in symbiotic stars. While our 
measured $\lambda7088/\lambda6830$ intensity ratio in \object{NGC~7027} may suffer from 
large uncertainties because of the weakness of the $\lambda$7088 
feature, its small value seems to suggest that the environs from which
these Raman lines arise in \object{NGC~7027} may differ from those in symbiotic
systems. In symbiotic binaries, the ultraviolet \ion{O}{vi} doublet emission from
the vicinity of a very hot white dwarf is Raman scattered in the
\ion{H}{i} atmosphere of a giant star companion. The nucleus of 
\object{NGC~7027} is not
known as a binary star. By analogy with the \ion{He}{ii} Raman lines observed
by \citet{pequignotbm,pequignot2003}, the \ion{O}{vi} Raman lines may be
formed in the photodissociation region (PDR) that corresponds to the
interface between the H$^+$ region and the large molecular envelope of
\object{NGC~7027}. The relatively small column density of this PDR, compared to the
atmosphere of a giant, may be compensated by the large covering factor.
The theoretical ratio of the \ion{O}{vi} $\lambda\lambda1032,1038$ lines is 
2 and the ratio of the
Raman cross sections is about 3.3. In a small optical depth approximation,
the intensity ratio of the Raman components should be of order 6.6, in
agreement with the observed intensities. The smaller ratio observed in
symbiotics may reflect a departure from the 2:1 ratio of the \ion{O}{vi} lines
(a similar departure is observed for \ion{C}{iv} and \ion{N}{v} in these objects;
see Schmid 1989). In \object{NGC~7027}, the red \ion{O}{vi} Raman lines may
allow one to indirectly determine the intensity of the UV \ion{O}{vi} 
lines,
which presumably arise from the high ionization region of the PN,
and thus provide a new useful constraint on the properties of the nucleus.

\setcounter{figure}{1}
\begin{figure}
\centering
\epsfig{file=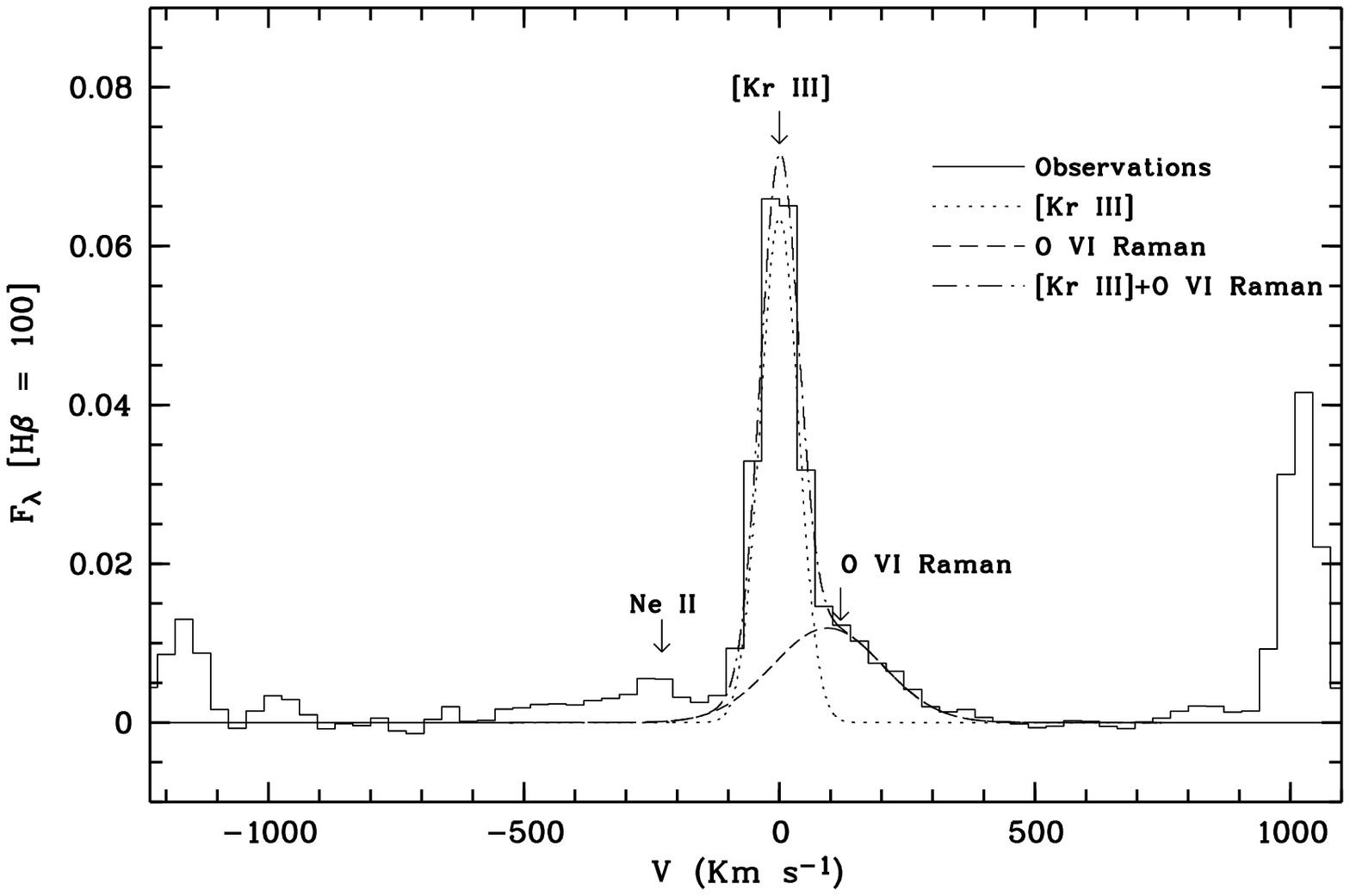,
height=6cm, bbllx=38, bblly=301, bburx=538, bbury=642, clip=, angle=0}
\caption{Detailed spectrum showing the region centered at 6828{\AA}. The
histogram is the observed spectrum.  The dotted-dashed line represents a
two-Gaussian fit, with the narrow and broad components represented by dotted
and dashed lines respectively. The broad feature redwards of the 
[\ion{Kr}{iii}] line is identified as the \ion{O}{vi} Raman line (see 
text).
}
\label{raman}
\end{figure}

Raman features have also been detected in the spectrum of the PN 
\object{NGC\,6302}
\citep{groves}. Both \object{NGC\,7027} and \object{NGC\,6302} have a very high
excitation class and show strong molecular emission. They thus may share some
common evolutionary properties.

\section{Reddening summary} \label{redden}

It is generally believed that a significant amount of dust coexists with the
ionized and neutral gas in \object{NGC\,7027} \citep[see, 
e.g.,][]{woodward}.  Large
variations of extinction across the nebula, caused by local dust, have been
detected by optical and radio imaging \citep{walton}. It follows that emission
lines arising from different ionized zones may suffer different amounts of
reddening. A comprehensive treatment of the dust extinction for 
the emission lines observed from NGC\,7027 requires detailed 
photoionization modeling and an
accurate treatment of the dust component (its spatial and size distributions,
chemical composition etc.), which is beyond the scope of the current work.
Efforts to correct for the effects of dust extinction on observed line fluxes
assuming a simplified geometry have been attempted previously by
\citet{seaton} and \citet{middlemass}.

On the other hand, our previous study shows that it is still a good
approximation to use the standard Galactic extinction curve for the 
diffuse
interstellar medium (ISM) to deredden the integrated spectrum of
\object{NGC\,7027} \citep{zhang}. Accordingly, we have dereddened all line 
fluxes by
\begin{equation} 
I(\lambda)=10^{c({\rm H}\beta)f(\lambda)}F(\lambda),
\end{equation} 
where $f(\lambda)$ is the standard Galactic extinction curve for
a total-to-selective extinction ratio of $R=3.2$ \citep{howarth}, and $c({\rm
H}\beta)$ is the logarithmic extinction at H$\beta$. 

\begin{figure}
\centering
\epsfig{file=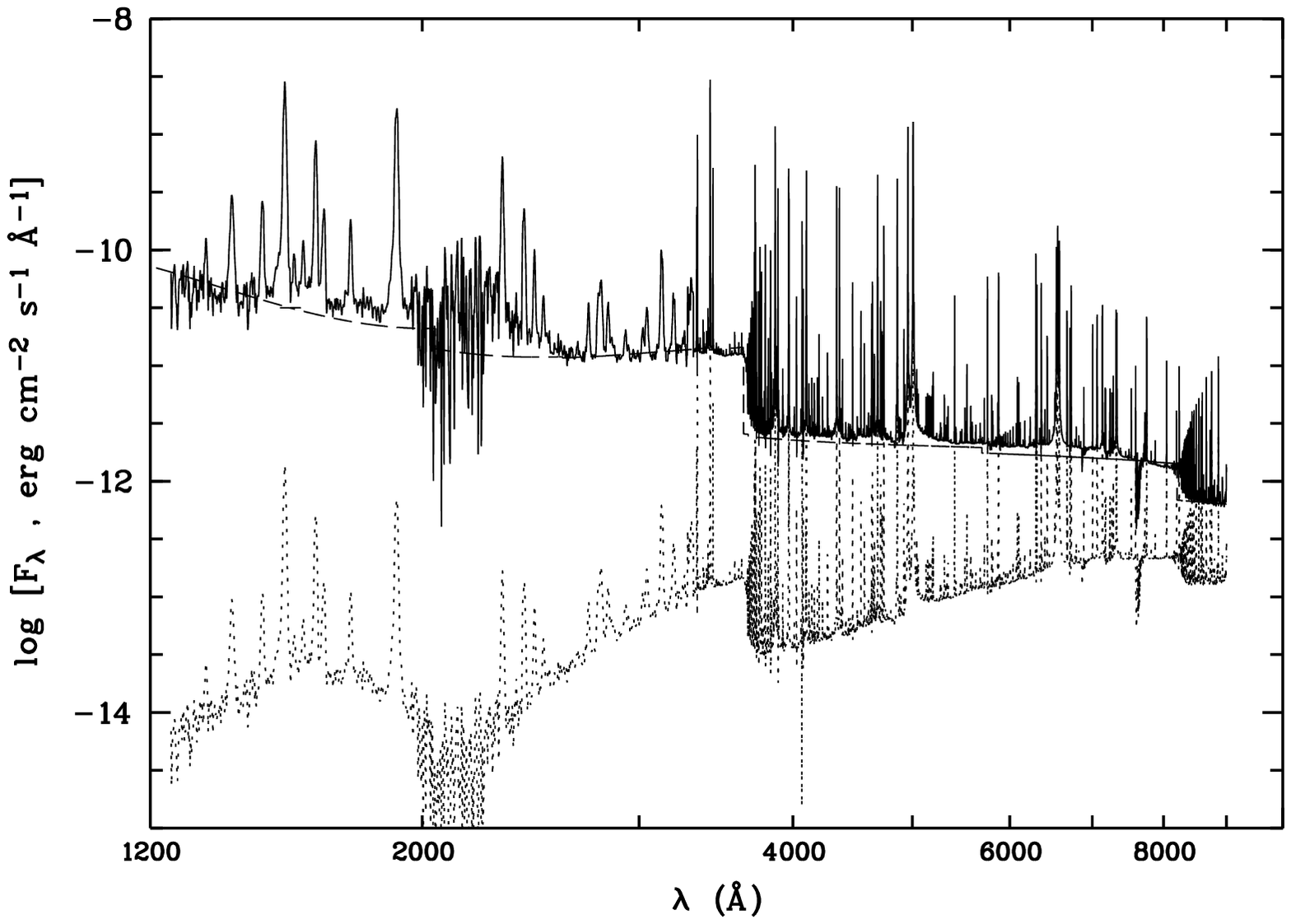,
height=6cm, bbllx=62, bblly=110, bburx=523, bbury=436, clip=, angle=0}
\caption{The UV and optical spectrum showing the 2200\,{\AA} extinction 
bump which
can be used to estimate total reddening towards NGC\,7027. The solid and dotted
lines show respectively the observed spectrum and that dereddened assuming
$c({\rm H}\beta)=1.44$.  The dashed line is a synthesized theoretical spectrum
of recombination line and continuum emission from ionized hydrogen and helium.}
\label{extiu}
\end{figure}

From the observed Balmer H$\alpha$/H$\beta$ and H$\gamma$/H$\beta$ ratios, we
deduce an average value of $c({\rm H}\beta)=1.34\pm0.02$. For $T_{\rm
e}=12\,800$\,K, ${\rm He}^{2+}/{\rm H}^{+}=0.041$ and ${\rm He}^{+}/{\rm
H}^{+}=0.058$ (see below), the 5-GHz free-free radio continuum flux density,
$S(5\,{\rm GHz})=6.921$\,Jy, combined with the total H$\beta$ flux, $\log
F({\rm H}\beta)=-10.12$\,(erg\,cm$^{-2}$\,s$^{-1}$) \citep{cahn}, yield $c({\rm
H}\beta)=1.37$, consistent with the value derived from the Balmer decrement
within the uncertainties. The observed \ion{He}{ii} $\lambda$1640/$\lambda$4686
ratio gives a slightly higher value, $c({\rm H}\beta)=1.42$.  As an alternative
way to determine $c$, we plot in Fig.~\ref{extiu} a synthesized spectrum of
NGC\,7027, computed taking into account contributions from the CS and from the
free-free and free-bound emission of ionized hydrogen and helium
(c.f. Zhang el al. 2004 for details).  By comparing the synthesized spectrum with
the observed one dereddened with a varying value of $c$, especially around the
2200\,{\AA} region of the UV extinction bump, as shown in Fig.~\ref{extiu}, we
derive $c({\rm H}\beta)=1.44$.  Although the \ion{He}{ii}
$\lambda$1640/$\lambda$4686 ratio and the dip at 2200\,{\AA} yield slightly
higher extinction constants compared to the values given by the Balmer
decrement and by the radio continuum flux density, presumably caused by the
presence of small dust grains within the nebula, we do not regard this as
significant and opt to use $c({\rm H}\beta)=1.37$ to deredden all measurements
from the UV to the IR.

\section{Physical conditions}

\subsection{CEL diagnostics}

\begin{figure}
\centering
\epsfig{file=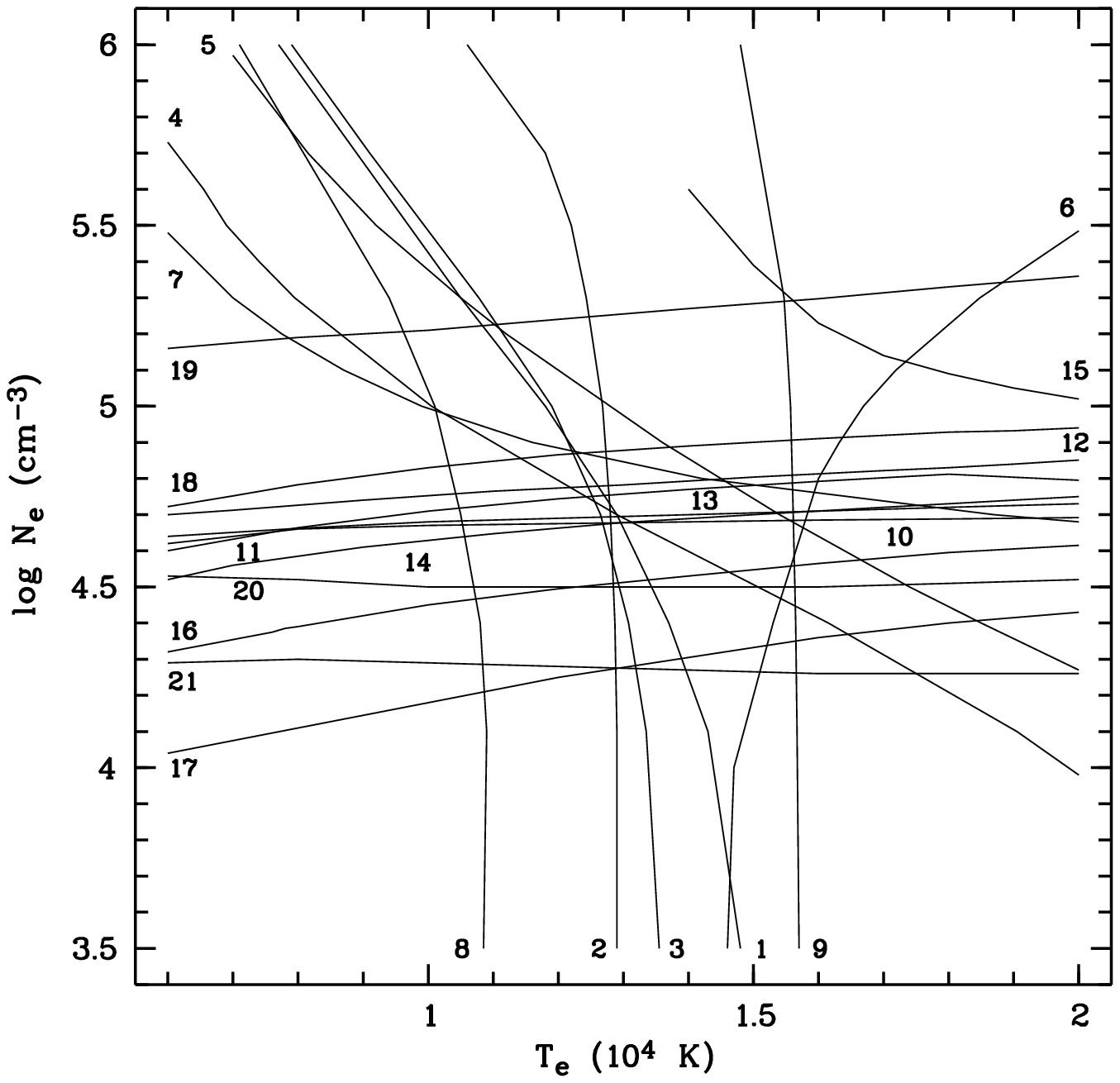, height=7.6cm,
bbllx=121, bblly=275, bburx=514, bbury=657, clip=, angle=0}
\caption{Plasma diagnostic diagram. Each curve is labelled with
an ID number given in Table~\ref{diagnostic}. }
\label{tnplane}
\end{figure}

The spectrum of \object{NGC\,7027} reveals a large number of CELs, useful for
nebular electron density and temperature diagnostics and abundance
determinations. Adopting atomic data from the references given in 
Appendix~B and
solving the level populations for multilevel ($\geq 5$) atomic models, we have
determined electron temperatures and densities from a variety of CEL ratios and
list the results in Table~\ref{diagnostic}.  Electron temperatures were derived
assuming a constant electron density of $\log N_{\rm e}=4.67$\,(cm$^{-3}$), the
average value yielded by a number of density-sensitive diagnostic ratios.
Likewise, an average electron temperature  of 14\,000\,K was assumed when
determining electron densities.  A plasma diagnostic diagram based on CEL
ratios is plotted in Fig.~\ref{tnplane}.

\setcounter{table}{4}
\begin{table}
\centering
\caption{Plasma diagnostics. \label{diagnostic}}
\begin{tabular}{clcc}
\hline\hline
{ID}           & {Diagnostic}     &
{Result}\\
\hline
& & $T_{\rm e}$\,(K)\\
 1&[Ne~{\sc III}] 15$\mu$m/($\lambda3868+\lambda3967$)&12900\\
 2&[Ar~{\sc iii}] $\lambda$7135/$\lambda$5192 &12800\\
 3&[O~{\sc iii}] ($\lambda4959+\lambda5007$)/$\lambda$4363 &12600\\
 4&[N~{\sc ii}] ($\lambda6548+\lambda6584$)/$\lambda$5754&12900\\
 5&[Ne~{\sc iv}] $\lambda$1602/($\lambda2422+\lambda2425$)&15300\\
 6&[O~{\sc iii}] $\lambda$4363/$\lambda$1663&14700\\
 7&[O~{\sc ii}] ($\lambda7320+\lambda7330$)/$\lambda$3726&18700\\
 8&[O~{\sc i}] $\lambda$5577/($\lambda6300+\lambda6363$)  &10400\\
 9&[Mg~{\sc v}] $\lambda$2783/5.6$\mu$m &15600\\
  & He~{\sc i} $\lambda6678/\lambda4471 $ &9430\\
  & He~{\sc i} $\lambda6678/\lambda5876 $ &12700\\
 & He~{\sc i} $\lambda7281/\lambda5876 $ &8800\\
  & He~{\sc i} $\lambda7281/\lambda6678 $ &10530\\
  & BJ/H\,11 &  12800\\
\noalign{\vskip3pt}
 & & $\log N_{\rm e}$\,(cm$^{-3}$)\\
 10&[Ar~{\sc iv}] $\lambda$4740/$\lambda$4711 & 4.77\\
 11&[C~{\sc iii}] $\lambda$1906/$\lambda$1909 & 4.80\\
 12&[Cl~{\sc iii}] $\lambda$5537/$\lambda$5517 &4.70\\
 13&[Ne~{\sc iv}] $\lambda$2425/$\lambda$2423 &4.69\\
 14&[Si~{\sc iii}] $\lambda$1884/$\lambda$1892 & 4.54\\
 15&[S~{\sc ii}] $\lambda$6731/$\lambda$6716  & $\ga5$\\
 16&[Ne~{\sc v}] 24$\mu$m/14$\mu$m  & 4.31\\
 17&[Ne~{\sc iii}] 15$\mu$m/36$\mu$m  & 4.89\\
 18&[O~{\sc ii}] $\lambda$3726/$\lambda$3729  & 4.68\\
 19&[Fe~{\sc iii}] $\lambda$4881/$\lambda$4701  & 5.27\\
 20&[Fe~{\sc iii}] $\lambda$4701/$\lambda$4733  & 4.50\\
 21&[Fe~{\sc iii}] $\lambda$4733/$\lambda$4754  & 4.27\\
   &H I Balmer decrement &   $\sim5$\\
   &He II Pfund decrement &  $\sim5$\\
\hline\end{tabular}
\end{table}

\citet{liubarlow} showed that the [\ion{N}{ii}] $\lambda5754$ line can be
enhanced by recombination excitation, leading to an overestimated electron
temperature being determined from the [\ion{N}{ii}]
$(\lambda6548+\lambda6584)/\lambda5754$ ratio. Using the formula given by
\citet{liubarlow}, we find that for this particular PN, the contribution by
recombination to the intensity of the [\ion{N}{ii}] $\lambda5754$ line is
negligible ($<1\%$).  We also find the presence of a weak \ion{He}{ii}
Raman line at 6546\,{\AA} hardly affects our estimate of the [\ion{N}{ii}]
$\lambda6548$ line flux -- our measurements yield a [\ion{N}{ii}]
$\lambda6548/\lambda6584$ ratio which is in excellent agreement with the
theoretical value of 2.9.

The electron temperature derived from the [\ion{O}{iii}]
$\lambda$4363/$\lambda$1663 ratio is 2100\,K higher than the value deduced from
the [\ion{O}{iii}] $(\lambda4959+\lambda5007)/\lambda4363$ ratio.  This may be
due to measurement uncertainties in the \ion{O}{iii}] $\lambda$1663 line flux.
The [\ion{O}{ii}] nebular to auroral line ratio,
$(\lambda3726+\lambda3729)/(\lambda7320+\lambda7330)$, yields an abnormally
high temperature of 18\,700\,K, 5800\,K higher than that determined from the
[\ion{N}{ii}] $(\lambda6548+\lambda6584)/\lambda5754$ ratio.  Similar
discrepancies have also been found in \object{M\,1-42}, \object{M\,2-36}
\citep{liuluo} and \object{NGC\,6153} \citep{liubarlow}. This is most likely
caused by the presence of high density clumps  in the nebula. The [\ion{O}{ii}]
$\lambda\lambda3726,3729$ nebular lines have much lower critical
densities ($\sim10^3$\,cm$^{-3}$) than the [\ion{O}{ii}]
$\lambda\lambda7320,7330$ auroral lines ($\sim10^6$\,cm$^{-3}$), and are thus
susceptible to suppression by collisional de-excitation in dense regions,
leading to an apparently high
$(\lambda3726+\lambda3729)/(\lambda7320+\lambda7330)$ temperature.

The line ratios of highly ionized species, [\ion{Ne}{iv}]
$\lambda$1602/($\lambda2422+\lambda2425$) and [\ion{Mg}{v}]
$\lambda$2783/5.6\,$\mu$m, yield  high electron temperatures of 15\,300\,K and
15\,600\,K, respectively.  For comparison, the line ratio of neutral species,
[\ion{O}{i}] $\lambda$5577/($\lambda6300+\lambda6363$), yields a much lower
value of 10\,400\,K. In addition, the [\ion{C}{i}] nebular to auroral line
ratio $(\lambda9824+\lambda9850)/\lambda8727$ gives a
temperature of only 8100~K for the warm transition regions around the nebula
\citep{liu1996}.
Our results thus suggest a negative temperature gradient across the nebula.  A
similar trend was previously found by \citet{salas}.

All density diagnostic ratios analyzed here yield compatible results. The only
exception is the [\ion{S}{ii}] $\lambda6731/\lambda6716$ doublet ratio, which
falls outside the high density limit for $T_{\rm e} < 14\,000$\,K.  The
[\ion{S}{ii}] $\lambda$6731 and $\lambda$6716 lines have relatively low critical
densities, $\log N_{\rm crit}=3.62$ and 3.21\,(cm$^{-3}$), respectively, at
$T_{\rm e} = 10\,000$\,K. The doublet is thus not suitable for density
determination for such a compact nebula.  Nevertheless, Fig.~\ref{tnplane}
shows that the observed [\ion{S}{ii}] $\lambda6731/\lambda6716$ ratio implies a
density higher than $10^5$\,cm$^{-3}$, which is at least a factor of two
higher than that derived from the [\ion{O}{ii}] $\lambda3726/\lambda3729$
doublet ratio. \citet{stanghellini} have found that the [\ion{S}{ii}] densities
of PNe are systematically higher than the corresponding values derived from the
[\ion{O}{ii}] lines.  \citet{rubin} ascribes this to the effects from a
dynamical plow by the ionization front, since S$^0$ has a lower ionization
potential than O$^0$ and thus the S$^+$ zone is closer to the H$^+$ edge.
Alternatively, this may also be caused by uncertainties in the atomic parameters
\citep{copetti,wang}. Note that \citet{liubar2001} find that densities derived
from the low critical density [\ion{O}{iii}] $88\mu{\rm m}/52\mu{\rm m}$ ratio
are systematically lower than given by high critical density optical diagnostic
ratios, such as the [\ion{Cl}{iii}] and [\ion{Ar}{iv}] doublet ratios, indicating
that density inhomogeneities are ubiquitous amongst PNe.

The [\ion{Fe}{iii}] $\lambda4881/\lambda$4701 ratio yields a relatively high
electron density of $\log N_{\rm e}=5.27$\,(cm$^{-3}$).  We note that the
\ion{N}{iii} V9 $\lambda4882$ line is blended with the [\ion{Fe}{iii}]
$\lambda4881$ line.  However, assuming \ion{N}{iii}
$I(\lambda4881)/I(\lambda4884)=0.05$, we find that the contribution of the
\ion{N}{iii} line to the $\lambda4881$ feature is negligible
($\sim0.2\%$).  On the other hand, the flux of the [\ion{Fe}{iii}] 
$\lambda4881$ line
could also be underestimated due to the possible effects of absorption by the
diffuse interstellar band (DIB) at 4882.56\,{\AA} \citep{tuairisg}, as
suggested by our detections in the spectrum of NGC\,7027 of DIBs at 5705, 
5780,
6204 and 6284\,{\AA} (see Fig.~\ref{all}). An underestimated [\ion{Fe}{iii}]
$\lambda4881$ line flux will lead to an overestimated [\ion{Fe}{iii}]
$\lambda4881/\lambda$4701 density.

\subsection{ORL diagnostics}

Table~\ref{diagnostic} gives the Balmer jump temperature, derived from the
ratio of the nebular continuum Balmer discontinuity at 3646\,{\AA} to H\,11
$\lambda3770$ using the following equation \citep[][]{liuluo}
\begin{equation}
T_{\rm e}({\rm BJ})=368\times(1+0.259Y^++3.409Y^{++})\times(\frac{{\rm BJ}}{\rm
H\,11})^{-1.5}\,{\rm K,}
\end{equation}
where BJ/H\,11 is in units of \AA$^{-1}$ and $Y^+$ and $Y^{++}$ are
He$^+$/H$^+$ and He$^{++}$/H$^+$ abundance ratios, respectively. Using
He$^+$/H$^+$ and He$^{++}$/H$^+$  ratios determined from \ion{He}{i} and 
\ion{He}{ii}
recombination lines (see next section), we obtain a $T_{\rm e}$(BJ) of
12\,800\,K. 

\begin{figure}
\centering
\epsfig{file=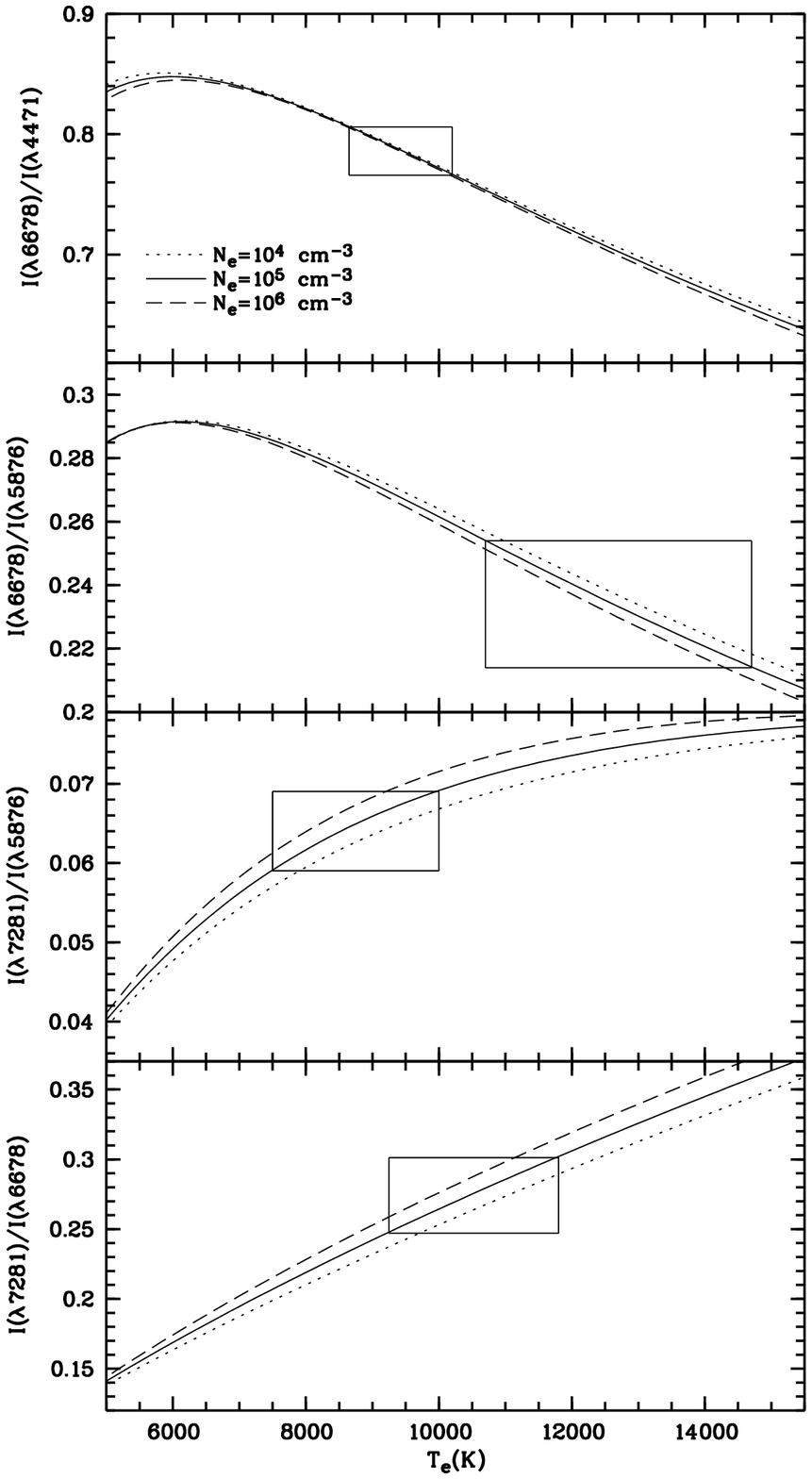, height=15.0cm,
bbllx=109, bblly=110, bburx=470, bbury=722, clip=, angle=0}
\caption{The \ion{He}{i}
$\lambda6678/\lambda4471$, $\lambda6678/\lambda5876$,
$\lambda7281/\lambda5876$ and $\lambda7281/\lambda6678$ ratios
as a function of electron temperature for $\log N_{\rm e}=4$ (dotted line),
5 (solid line) and 6 (dashed line) (cm$^{-3}$).
The boxes are the observed values with measurement uncertainties.}
\label{heidia}
\end{figure}

We have also used the \ion{He}{i} line ratios to determine the average
\ion{He}{i} line emission electron temperature.  In Fig.~\ref{heidia}, we plot
\ion{He}{i} $\lambda6678/\lambda4471$, $\lambda6678/\lambda5876$,
$\lambda7281/\lambda5876$ and $\lambda7281/\lambda6678$ ratios as a function of
electron temperature. The measured line ratios along with their uncertainties
are overplotted.  The \ion{He}{i} line emission coefficients used here are from
\citet{benjamin} under the assumption of Case B recombination.  An electron
density of $N_{\rm e}=10^5$\,cm$^{-3}$, as deduced by the \ion{H}{i} Balmer
decrement and the \ion{He}{ii} Pfund decrement (see below), was assumed when
determining temperatures, although the results are insensitive to the adopted
electron density, as shown in Fig.~\ref{heidia}.  The results are presented in
Table~\ref{diagnostic}.  Fig.~\ref{heidia} shows that the temperatures derived 
from the \ion{He}{i} line ratios range between 7500 and 14700~K.
Considering certain advantages over other \ion{He}{i} ratios, as discussed 
in \citet{zhang2005}, the \ion{He}{i} $\lambda7281/\lambda6678$ ratio gives 
the most reliable result, which is only 1.5$\sigma$ lower than the value 
derived from the [\ion{O}{iii}] forbidden line ratio.

\begin{figure*}
\centering
\epsfig{file=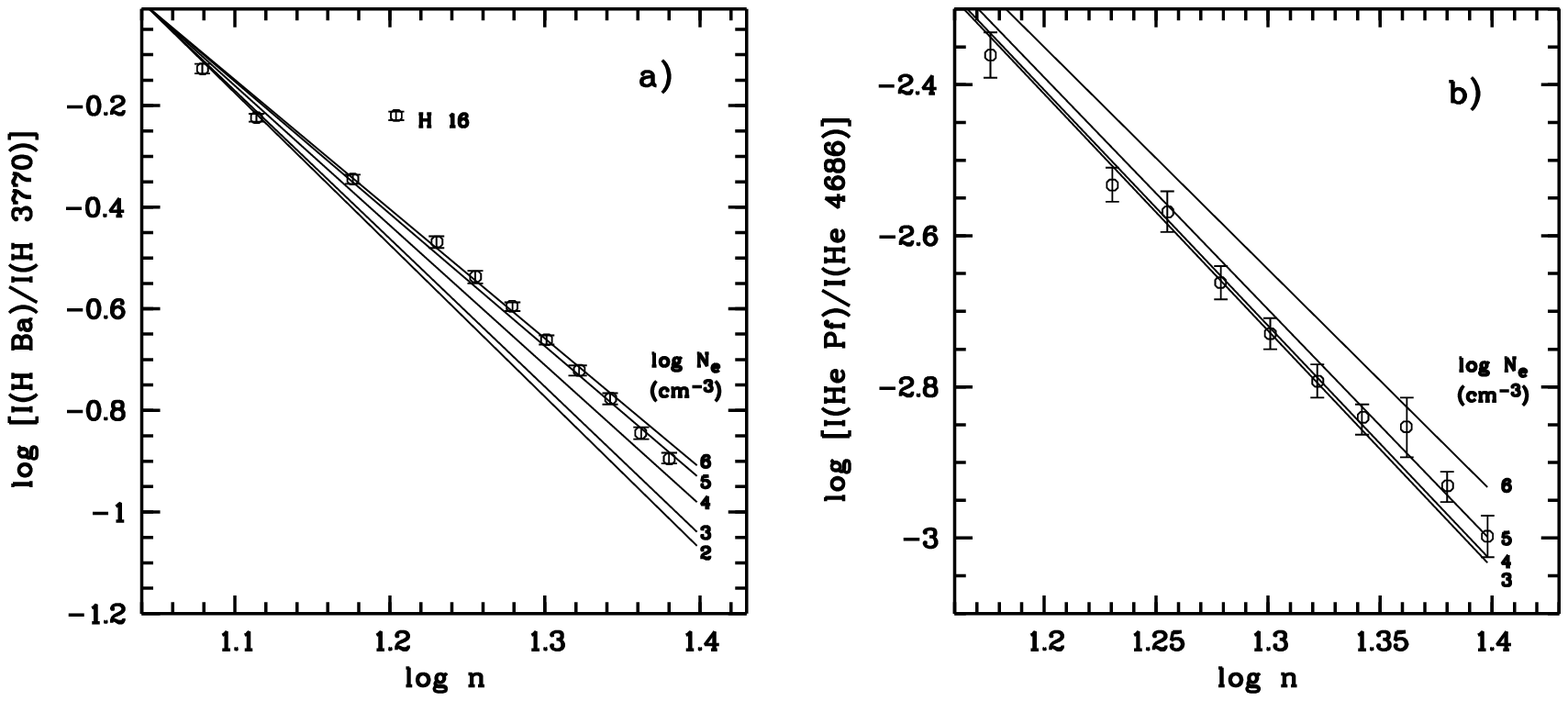, height=8cm,
bbllx=41, bblly=425, bburx=533, bbury=649, clip=, angle=0}
\caption{Intensity ratios of a) \ion{H}{i} Balmer lines to H~11 $\lambda$3770
and b) \ion{He}{ii} Pfund lines to \ion{He}{ii} $\lambda$4686, as a function of
the principal quantum number $n$ of the upper level of the transition. H~16 at
3703.86\,{\AA} is blended with the \ion{He}{i} $\lambda3705.12$ line. Solid
lines are theoretical values for electron densities of $N_{\rm e}=10^2$ to
10$^6$\,cm$^{-3}$ at a constant electron temperature of 12\,800\,K.}
\label{re_ne}
\end{figure*}

Electron densities have been derived from the \ion{H}{i} Balmer decrement
and the \ion{He}{ii} Pfund decrement.  Fig.~\ref{re_ne} shows the observed
intensity ratio of the high-order \ion{H}{i} Balmer lines ($n\rightarrow2$,
$n=12,13,...,24$) to H~11 $\lambda3770$ and the \ion{He}{ii} Pfund lines
($n\rightarrow5$, $n=15,17,...,25$) to \ion{He}{ii} $\lambda$4686 as a function
of the principal quantum number $n$ of the upper level. Theoretical intensity
ratios for different electron densities are overplotted assuming an electron
temperature of 12\,800\,K, as deduced from the \ion{H}{i} Balmer discontinuity.
As shown in Fig.~\ref{re_ne}, both the \ion{H}{i} Balmer decrement and the
\ion{He}{ii} Pfund decrement yield a best fit at $N_{\rm e}\sim10^5$\,cm$^{-3}$,
in reasonable agreement with densities derived from CEL diagnostics.  The
result rules out the possibility that \ion{H}{i} and \ion{He}{ii} lines arise
from dense regions with an electron density in excess of $10^7$\,cm$^{-3}$ as
proposed by \citet{kaler}.  In principle the \ion{H}{i} Paschen decrement and
the \ion{He}{ii} Pickering decrement may also be used to probe electron density
\citep{pequignotba}.
However, both series are strongly affected by telluric absorption and line
blending, and thus are omitted in the current analysis.

\section{Elemental abundances}

\subsection{Ionic abundances from CELs}

Ionic abundances are derived from CELs by
\begin{equation}
\frac{N({\rm X}^{i+})}{N({\rm H}^+)}=\frac{I_{jk}}{I_{{\rm H}\beta}}\frac{\lambda_{jk}}{\lambda_{{\rm H}\beta}}\frac{\alpha_{{\rm H}\beta}}{A_{jk}}\left[{\frac{N_j}{N({\rm X}^{i+})}}\right]^{-1}N_{\rm e},
\end{equation}
where $I_{jk}/I_{{\rm H}\beta}$ is the intensity ratio of the ionic line to
H$\beta$, $\lambda_{jk}/\lambda_{{\rm H}\beta}$ is the wavelength ratio of the
ionic line to H$\beta$, $\alpha_{{\rm H}\beta}$ is the effective recombination
coefficient for ${\rm H}\beta$, $A_{jk}$ is the Einstein spontaneous transition
rate of the ionic line, and $N_j/N({\rm X}^{i+})$ is the fractional population
of the upper level where the ionic line arises and is a function of electron
temperature and density.

Ionic abundances derived from UV, optical and infrared CELs are presented in
Table~\ref{cel_ab}. Based on plasma diagnostic results presented in the
previous section, a constant electron density of $\log N_{\rm
e}=4.67$\,cm$^{-3}$ was assumed for all ionic species.  Electron temperatures
of 12\,600\,K and 15\,500\,K were assumed throughout for ions with ionization
potentials lower and higher than 50\,eV, respectively.  In 
Table~\ref{cel_ab},
the `adopted' denotes average and adopted values derived from individual lines
weighted by intensity.  

Note that Mg$^+$ and Fe$^+$ exist mainly in the 
PDR outside the ionized zone (defined by H$^+$), given the very low
($<8$\,eV) ionization potentials of neutral magnesium and iron.  Thus Mg$^+$
and Fe$^+$ ionic abundances are not listed in Table~\ref{cel_ab}, even though
several \ion{Mg}{ii} and [\ion{Fe}{ii}] CELs have been detected.  Similarly, the
\ion{C}{ii} 158-$\mu$m line should also arise mainly from the PDR, rather than
the ionized region. In fact, the C$^+$/H$^+$ ionic abundance ratio derived
from the far-IR fine-structure line is a factor of six higher than the value
derived from the UV CEL $\lambda$2324. The large discrepancy remains even if one
adopts the lower 158-$\mu$m line flux given by \citet{liuiso}, rather than the
more recent flux presented in \citet{liubar2001} based on a more recent version
of the LWS data processing pipeline. 

\subsection{Ionic abundances from ORLs}

From measured intensities of ORLs, ionic abundances can be derived using
the following equation
\begin{equation}
\frac{N({\rm X}^{i+})}{N({\rm H}^+)}=\frac{I_{jk}}{I_{{\rm H}\beta}}\frac{\lambda_{jk}}{\lambda_{{\rm H}\beta}}\frac{\alpha_{{\rm H}\beta}}{\alpha_{jk}},
\end{equation}
where $\alpha_{jk}$ is the effective recombination coefficient for the ionic
ORL and is taken from the references listed in Appendix~B.  Ionic abundances
derived from ORLs depend only weakly on the adopted temperature ($\sim T_{\rm
e}^\alpha$, $|\alpha|<1$), and are essentially independent of $N_{\rm e}$. A
constant temperature of $T_{\rm e}=12\,800$\,K, as deduced from the \ion{H}{i}
Balmer jump, was assumed throughout. Ionic abundances derived from ORLs are
presented in Table~\ref{orl_ab}.

\subsection{Comparison of ionic abundances derived from CELs and ORLs}

\begin{figure*}
\centering
\epsfig{file=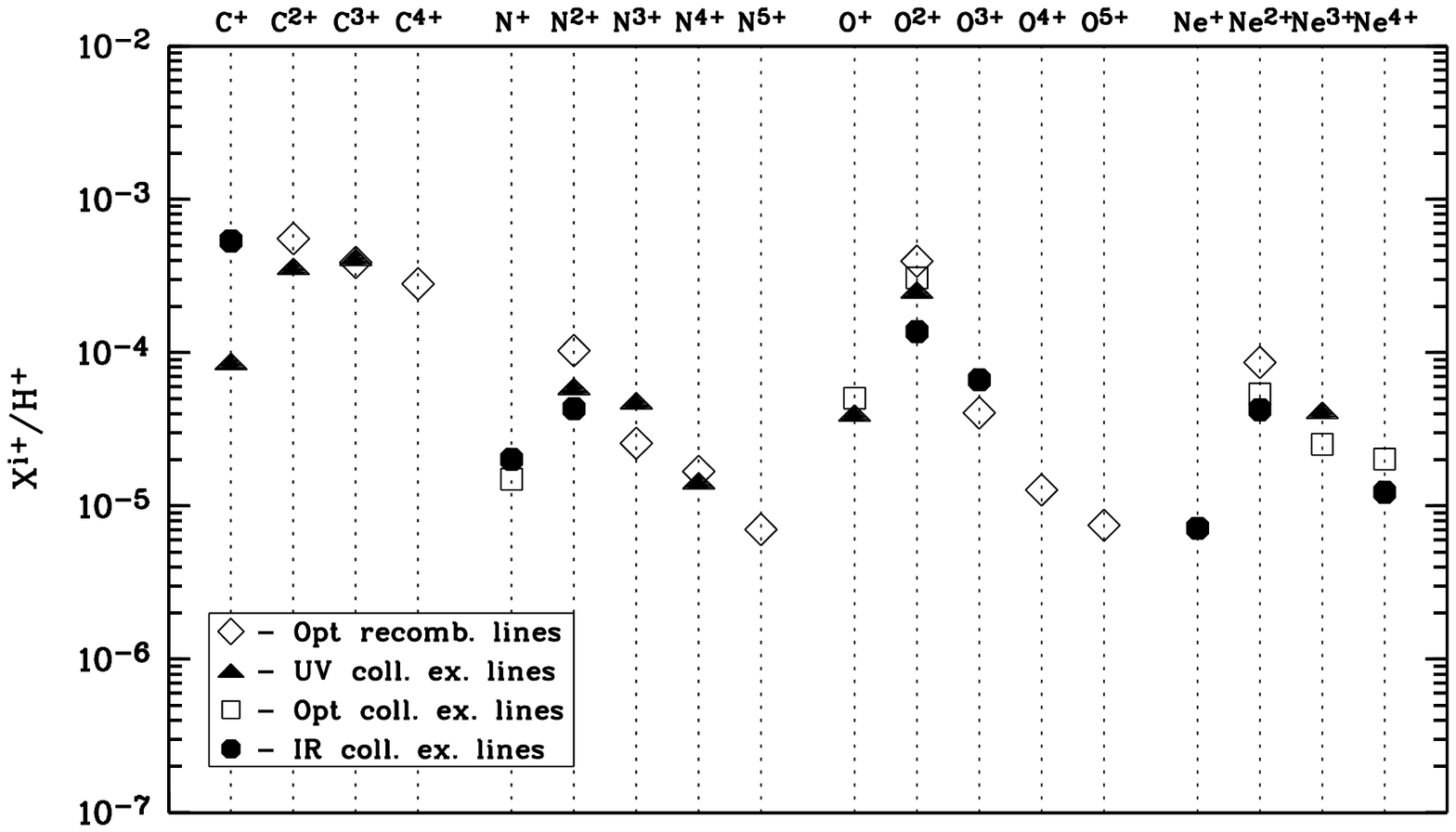, height=9.0cm,
bbllx=62, bblly=195, bburx=547, bbury=469, clip=, angle=0}
\caption{Comparison of C, N, O and Ne ionic abundances derived from ORLs
with those from UV, optical and IR CELs.}
\label{com_ab}
\end{figure*}

In Table~\ref{comp} and Fig.~\ref{com_ab} we compare C, N, O and Ne ionic abundances derived from
CELs with those derived from ORLs. Table~\ref{comp} and Fig.~\ref{com_ab} 
show that the ionic
abundances of doubly ionized species derived from ORLs, namely those of
C$^{2+}$/H$^+$, N$^{2+}$/H$^+$, O$^{2+}$/H$^+$ and Ne$^{2+}$/H$^+$, are 
only
slightly higher that those derived from CELs, with an abundance discrepancy
factor of about 1.5.

\setcounter{table}{7}
\begin{table}
\centering
\caption{Comparison of ionic abundances. \label{comp}}
\begin{tabular}{lccc}
\hline\hline
Ion &\multicolumn{2}{c}{N(X$^{i+}$)/N(H$^{i+}$)} &{ORL/CEL}\\ 
\cline{2-3}
& {CEL}     &
{ORL} & \\
\hline
C$^{2+}$& 3.85$\times$10$^{-4}$& 5.54$\times$10$^{-4}$ & 1.4 \\
C$^{3+}$& 4.18$\times$10$^{-4}$& 3.87$\times$10$^{-4}$ & 0.9 \\
N$^{2+}$& 5.95$\times$10$^{-5}$& 1.03$\times$10$^{-4}$ & 1.7 \\
N$^{3+}$& 4.85$\times$10$^{-5}$& 2.56$\times$10$^{-5}$ & 0.5 \\
N$^{4+}$& 1.46$\times$10$^{-5}$& 1.68$\times$10$^{-5}$ & 1.2 \\
O$^{2+}$& 3.06$\times$10$^{-4}$& 3.95$\times$10$^{-4}$ & 1.3 \\
O$^{3+}$& 6.63$\times$10$^{-5}$& 4.04$\times$10$^{-5}$ & 0.6 \\
Ne$^{2+}$&5.35$\times$10$^{-5}$& 8.63$\times$10$^{-5}$ & 1.6 \\
\hline\end{tabular}
\end{table}

The dichotomy between the ORL and the CEL abundances is an open problem in
nebular astrophysics (see Liu 2003 and references therein).  A chemically
homogeneous nebula with temperature and density variations fails to account for
the discrepancy \citep[c.f.][]{liu}.  Instead detailed multi-waveband analyses
of a large sample of PNe point to the presence of H-deficient inclusions
embedded in the diffuse nebular gas
\citep{tsamis03,tsamis04,liuya,liuyb,wessonr}. A two-abundance model, first
proposed by \citet{liubarlow}, provides the most plausible explanation for this
problem.  The model predicts $T_{\rm e}$(\ion{He}{i})$\la T_{\rm e}$(BJ)$\la
T_{\rm e}$([\ion{O}{iii}]).  \citet{liuluo} found that the ORL/CEL abundance
discrepancy is positively correlated with the difference between $T_{\rm
e}$([\ion{O}{iii}]) and $T_{\rm e}$(BJ). It has also been found that the
discrepancy increases as the nebula expands and ages 
\citep{GarDin,tsamis04,zhangl}. \object{NGC\,7027} is a
relatively young and compact PN. The small ORL to CEL discrepancies, both 
for
temperature determinations and for abundance determinations, are therefore
consistent with the general picture found for other PNe.

Interestingly, it seems that compared to doubly ionized species, abundance
discrepancy factors for ionic species of even higher ionization degree, such as
C$^{3+}$, N$^{3+}$, N$^{4+}$ and O$^{3+}$, are even smaller. This result is
however arguable considering uncertainties in both measurements and in the
analysis. If real, then it may indicate that the postulated ultra-cold
H-deficient inclusions have a lower ionization degree compared to the diffuse
nebula of ``normal'' chemical composition.

\subsection{Total abundances}

Total elemental abundances derived for \object{NGC\,7027} are presented and
compared to values published in the literature in Table~\ref{abco}.  Except for
helium, all elemental abundances given in the table are based on CEL analyses.
The average abundances deduced for a large sample of Galactic PNe by
\citet{kingsburgh} and the solar photospheric abundances compiled by
\citet{lodders} are also given in the table.

\begin{table*}
\centering
\noindent \parbox{15cm}{
\caption{Elemental abundances in NGC\,7027, in units
such that $\log N({\rm H})=12.0$. \label{abco}}
\begin{tabular}{ccccccccc}
\hline\hline
{Element}         &
{This paper}           & {(1)}  &
{(2)}          & {(3)}    &
{(4)}  & {(5)} &{Average$^{\mathrm{a}}$}  &
Solar$^{\mathrm{b}}$\\
\hline
 He &11.00 &  11.03 & 11.00 & 11.05 & 11.02 & 11.00   &11.06   & 10.90\\
 C  &9.10  &  8.78  & 8.98  & 8.84  & 9.11  & 8.98    &8.74    &  8.39\\
 N  &8.14  &  8.20  & 8.21  & 8.10  & 8.28  &   &8.35    &  7.83\\
 O  &8.66  &  8.61  & 8.71  & 8.49  & 8.75  & 8.66    &8.68    &  8.69\\
 Ne &8.07  &  8.00  & 8.14  & 8.00  & 8.04  &   &8.09    &  7.87\\
 Mg &7.60  &  7.34  & & 7.41  & 7.33  &   &  &  7.55\\
 S  &6.92  &  6.97  & & 6.86  & 6.90  &   &6.92    &  7.19\\
 Cl &5.21  &  5.04  & & 5.30  & &   &  &  5.26\\
 Ar &6.30  &  6.36  & & 6.32  & 6.40  &   &6.39    &  6.55\\
 Fe &5.58  &  & & & 6.00  &   &  &  7.47\\
\hline\end{tabular}
\begin{list}{}{}
\item[$^{\mathrm{a}}$] Average abundances of Galactic PNe \citep{kingsburgh};
\item[$^{\mathrm{b}}$] \citet{lodders}.
\end{list}

Reference: { (1) Bernard-Salas et al. 2001; (2) Kwitter \& Henry 1996;
(3) Keyes et al. 1990; (4) Middlemass 1990; (5) Gruenwald \&
P{\' e}quignot 1989.}
}
\end{table*}

The helium elemental abundance is calculated from ${\rm He/H}={\rm
He^+/H^+}+{\rm He^{2+}/H^+}$.  For a high-excitation PN such as
\object{NGC\,7027}, very little neutral helium is expected to exist within the
ionized region, and can therefore be ignored.

Whenever available, ionization correction factors ($ICF$s) given by
\citet{kingsburgh} were used to compute the CEL elemental abundances.

The carbon abundance is calculated using the equation
\begin{equation}
\frac{{\rm C}}{{\rm H}}=ICF({\rm C})\times(\frac{{\rm C}^+}{{\rm H}^+}+\frac{{\rm C}^{2+}}{{\rm H}^+}+\frac{{\rm C}^{3+}}{{\rm H}^+}),
\end{equation}
where
\begin{eqnarray}
ICF({\rm C})&=&(1-\frac{2.7{\rm N}^{4+}}{{\rm N}^{+}+{\rm N}^{2+}+{\rm N}^{3+}+{\rm N}^{4+}})^{-1}\\
             & =&1.40.\nonumber\\
\nonumber
\end{eqnarray}

For oxygen, we use
\begin{equation}
\frac{{\rm O}}{{\rm H}}=ICF({\rm O})\times(\frac{{\rm O}^+}{{\rm H}^+}+\frac{{\rm O}^{2+}}{{\rm H}^+}+\frac{{\rm O}^{3+}}{{\rm H}^+}),
\end{equation}
where
\begin{eqnarray}
ICF({\rm O})&=&(1-\frac{0.95{\rm N}^{4+}}{{\rm N}^{+}+{\rm N}^{2+}+{\rm N}^{3+}+{\rm N}^{4+}})^{-1}\\
             & =&1.11.\nonumber\\
\nonumber
\end{eqnarray}

For sulphur, \citet{kingsburgh} give an $ICF$ formula for the case that only
S$^{+}$ and S$^{2+}$ are measured. For \object{NGC\,7027}, we have detected
lines from S$^{+}$, S$^{2+}$ and S$^{3+}$. Because of its high-excitation,
S$^{4+}$ and S$^{5+}$ are also expected to be present in a significant fraction
and thus need to be corrected for. From the photoionization model presented by
\citet{shields}, we estimate an $ICF$(S) of 2.13. However, the photoionization
model constructed by \citet{middlemass} gives a lower $ICF$(S) of 1.31.
Here we have opted for an average value of 1.72. 

\citet{kingsburgh} did not discuss magnesium and iron.  For magnesium, no CELs
are available to determine the ${\rm Mg}^{2+}/{\rm H}^{+}$ ratio.  On the hand,
\citet{barlow} show that whatever effect is enhancing the ORL abundances of
second-row elements, such as C, N, O and Ne, it seems that it does not affect
magnesium, the only third-row element that has been studied using an ORL.  Here
we have simply added the ${\rm Mg}^{2+}/{\rm H}^{+}$ ionic abundance ratio
derived from the \ion{Mg}{ii} $\lambda$4481 ORL to ${\rm Mg}^{3+}/{\rm H}^{+}$
and ${\rm Mg}^{4+}/{\rm H}^{+}$ derived from IR CELs to obtain the total Mg
elemental abundance. For iron, the only missing ionization stage that needs
to be corrected for is Fe$^{4+}$. According to the photoionization model of
\citet{shields}, we estimate an $ICF$(Fe) of 1.13.  Note that Mg$^+$ and Fe$^+$
exist predominantly in the PDRs outside the ionized region. Although
a small amount of these ions is present in the ionized region and
emits CELs, we neglect the ionic concentrations in Mg$^+$ and Fe$^+$ in 
the calculation of the total elemental abundances of magnesium and iron
in the ionized zone.

For nitrogen, neon, argon and chlorine, no ionization corrections are required
and the elemental abundances are given by simply adding up individual observed
ionic abundance ratios.

As can be seen from Table~\ref{abco}, elemental abundances derived from
individual analyses are generally in good agreement. The largest
discrepancy is for carbon, by a factor of 2.14. Except for \citet{salas} and
the current work, where an empirical method is used, all other abundance
analyses presented in Table~\ref{abco} were based on photoionization modeling.
Abundances deduced in the current work are within the range of values reported
in previous papers, except for Ar, Mg and Fe. For Ar, although our abundance is
slightly lower than values found by others, we deem the difference
insignificant. The Mg abundance obtained in the
current work is $74\%$ higher than the average value of previous studies. Given
that ${\rm Mg}^{2+}$ is the dominant ionic species of magnesium, owing to the
very large ionization potential span of doubly ionized magnesium, from 15\,eV
to 80\,eV, and the fact that in all the previous works the Mg$^{2+}$ abundance was
not determined and was corrected for using either an $ICF$ or photoionization
model method, we regard the  Mg abundance obtained in the current work as 
more
reliable.  Amongst the previous studies, only \citet{middlemass} gives 
an iron
abundance and his value is significantly higher than ours. \cite{perinotto}
recently calculated iron abundances in a number of PNe and find an Fe/H 
ratio of
 $4\times10^{-7}$ [5.60 in units such that $\log N({\rm H})=12$] for 
\object{NGC\,7027}, 
which is in excellent agreement with our result.

Our analysis yields a ${\rm C}/{\rm O}$ ratio of 2.72. This higher than 
solar C/O ratio indicates that \object{NGC\,7027} is carbon-enriched.  The 
N/O ratio is
0.5, about three times larger than the solar value. According to the criteria
defined by \citet{peimbert}, \object{NGC\,7027} is not helium and nitrogen rich
enough to be a Type-I PN. 
Magnesium is in good agreement with the solar value. The high magnesium
abundance suggests that depletion of magnesium onto dust grains is
insignificant in this young PN, contrary to the result of \citet{salas}. 
The Fe
abundance is well below the solar value, by a factor of about 85, 
suggesting
that an effective condensation of gas phase iron into dust grains has occurred
in this carbon-rich PN. 

\section{The central star}

It is well-known that \object{NGC\,7027} is a high-excitation PN. Based on the
excitation classification scheme proposed by \citet{dopita90}, which
makes use of the observed $I$([\ion{O}{iii}]$\lambda$5007)/$I({\rm H}\beta)$ line
ratio for low and intermediate excitation PNe (E.C. $\leq 5$), and the
$I$(\ion{He}{ii}$\lambda4686$)/$I({\rm H}\beta)$ ratio for higher excitation
class PNe, we find that \object{NGC\,7027} has a E.C. of 6.97 from the observed
$I$(\ion{He}{ii}$\lambda4686$)/$I({\rm H}\beta)$ ratio.

The complete spectral data from the UV to IR accumulated here allow us to
calculate the energy-balance temperature (the Stoy temperature) of the CS of
\object{NGC\,7027}. The method was first introduced by \citet{stoy} and then
advanced by \citet{kaler} and \cite{preite}. The method rests on the idea that
the total flux of all CELs (which provide the major cooling rate of the nebula)
relative to H$\beta$, is a measure of the average heating rate per
photoionization incident (dominated by photoionization of hydrogen given its
high abundance), and thus depends on the effective temperature of the central
star. Unlike the Zanstra method \citep{zanstra}, the Stoy temperature is
insensitive to the nebular optical depth to ionizing photons and is thus 
applicable to
optically thin or partly optically thin nebulae. The major drawback of this
method is that it requires the fluxes of all major CELs to be measured, which
poses a major observational challenge. For the current data set, essentially
all major CELs have been detected. For the remaining few unseen strong lines,
their fluxes can be estimated from the observed lines emitted by the same 
ion.
Even for the very few unseen ions, fluxes of CELs emitted by those ions can be
calculated once we have estimated the abundances of those ions using the
ionization correction method. According to our calculation, the total
contribution of all unseen CELs is less than 2\%. Using the Stoy method
developed by \citet{preite} and assuming blackbody radiation for the CS, we
derive that the CS has an effective temperature of 219\,000\,K. 

Adopting a distance 880\,pc to \object{NGC\,7027} \citep{masson}, we obtain a
luminosity of $8100\,L_\odot$ and a photospheric radius of $0.063\,R_\odot$ for
the CS. Then, from the evolutionary tracks for hydrogen-burning post-AGB 
stars
calculated by \citet{vassiliadis} and the approximate formula for the
initial-final mass relationship given by the same authors, we estimate that the
CS has a core mass $M_c=0.74\,M_\odot$ and an initial mass $M_i=3.2\,M_\odot$.
For comparison, \citet{salas} estimate an initial mass between 3--4\,$M_\odot$
by comparing the elemental abundances of \object{NGC\,7027} with the
semi-analytical models of \citet{marigo}.

The properties of the CS are summarized in Table~\ref{star}, and compared to
the values obtained by \citet{latter}. Note that \citet{latter} determined the
effective temperature using the Zanstra method and their value is only slightly
lower than ours, suggesting that \object{NGC\,7027} is largely optically thick,
consistent with the presence of an extensive neutral and molecular
envelope around the ionized region of this PN.

\begin{table}
\centering
\caption{Properties of the central star. \label{star}}
\begin{tabular}{lcc}
\hline\hline
{Parameter} & {Present paper} &
 {Latter et al.}\\
\hline
$T_*$\,(10$^4$K)& 21.9$^{\mathrm{a}}$ & 19.8$^{\mathrm{b}}$\\
$L_*$\,(10$^3$$L_\odot$)     & 8.10   & 7.71    \\
$R_*$\,(10$^{-2}$$R_\odot$)    & 6.3   &  7.5  \\
$M_c$\,($M_\odot$)     & 0.74    &  0.7   \\
$M_i$\,($M_\odot$)     & 3.2     &   4    \\
\hline\end{tabular}
\begin{list}{}{}
\item[$^{\mathrm{a}}$]Based on the Stoy method;
\item[$^{\mathrm{b}}$]Based on the Zanstra method.
\end{list}
\end{table}

\section{Summary}

In this paper, we have presented deep optical spectrum of the bright PN
\object{NGC\,7027} from 3310\,{\AA} to 9160\,{\AA}, obtained by uniformly
scanning a long slit across the nebula, thus yielding average line fluxes for
the whole nebula.  A total of 1174 line identifications have been obtained,
with intensities ranging from $10^{-5}$ to $10^2$ of ${\rm H}\beta$. The
comprehensive line list should prove useful for future spectroscopic
study of emission line nebulae.

The optical spectrum, together with the {\it IUE}\ spectra in the UV and the
{\it ISO}\ spectra in the IR, have been used to determine the physical
conditions and elemental abundances.  The results show that \object{NGC\,7027}
is characterized by an average electron temperature of 14\,000\,K and a density
of 47\,000\,cm$^{-3}$.  The temperatures derived from the [\ion{O}{iii}] 
forbidden
line ratio and from \ion{H}{i} and from \ion{He}{i} recombination line 
ratios are
in agreement within the errors.  
Our analysis also yields a negative temperature gradient as a function of
   nebular radius. In calculating the ionic abundances, we have used
   a simplified two-zone model, adopting, respectively, an electron 
   temperature of $T_{\rm e} = 12\,600$ and 15\,500\,K for ions with
   ionization potentials lower or higher than 50\,eV.
Our abundance analysis show that ORLs yield only
marginally higher metal abundances than CELs.  This is in accord with the
general pattern found for a large sample of PNe
\citep{liuluo,liuya,liuyb,tsamis03,tsamis04,wessonr,zhangl} that the ORL/CEL
abundance discrepancy increases with increasing difference between the
[\ion{O}{iii}] and the \ion{H}{i} Balmer jump temperatures and with decreasing
electron density.

The integrated nebular line spectra from the UV to the IR allowed us to 
determine precisely
the effective temperature of the ionizing CS using the Stoy, or 
enrgy-balance, method. We find a
high temperature of $T_{\rm eff} = 219\,000$\,K.  For the first time, the
Raman-scattered \ion{O}{vi} features at 6830 and 7088\,{\AA} have been detected
in this PN, suggesting that abundant neutral gas is present around
the ionized regions of this high-excitation PNe.

The spectral data set presented in the current paper, together
with extensive high resolution images in the optical and in the radio
\citep{robberto,bryce,bains}, should provide tight constraints for future
photoionization modeling of \object{NGC\,7027}.

\begin{acknowledgements}
We are grateful to Y.-R. Sun and Y. Liu for their help with the preparation of 
this paper. YZ acknowledges the award of an Institute Fellowship from STScI.
The work of YZ and XWL has been supported partially by Chinese NSFC Grant 
No.~10325312. We thank P. J. Storey for his comments. We are also
happy to thank the referee, M. Perinotto, for
comments which have helped us to improve the paper.
\end{acknowledgements}

\setcounter{table}{5}
\begin{longtable}{llcl}
\caption{Ionic abundances from CELs. \label{cel_ab}}\\
\hline\hline
{Ion} & {$\lambda$} & {$I(\lambda)$}&
{N(X$^{i+}$)/N(H$^+$)} \\
 & {({\AA})} &
{[$I({\rm H}\beta)=100$]} & \\
\hline
\endfirsthead
\caption{continued.}\\
\hline\hline
{Ion} & {$\lambda$} & {$I(\lambda)$}&
{N(X$^{i+}$)/N(H$^+$)} \\
 & {({\AA})} &
{[$I({\rm H}\beta)=100$]} & \\
\hline
\endhead
\hline
\endfoot
C$^{+}$& 2324&177$^{\mathrm{a}}$& 8.74$\times10^{-5}$\\
        &15.8$\mu$m&2.4&5.36$\times10^{-4}$\\
&adopted&& 8.74$\times10^{-5}$\\
C$^{2+}$&1908&726&3.85$\times10^{-4}$\\
C$^{3+}$&1550&1125&4.18$\times10^{-4}$\\
N$^+$&5754&5.76&1.60$\times10^{-5}$\\
    &6548&33.1&1.53$\times10^{-5}$\\
    &6584&93.6&1.47$\times10^{-5}$\\
    &122$\mu$m&0.028&2.01$\times10^{-5}$\\
&adopted&& 1.49$\times10^{-5}$\\
N$^{2+}$&1750&27.1&6.01$\times10^{-5}$\\
        &57$\mu$m&0.99&4.30$\times10^{-5}$\\
&adopted&& 5.95$\times10^{-5}$\\
N$^{3+}$&1485&42.3&4.85$\times10^{-5}$\\
N$^{4+}$&1240& 35.9&1.46$\times10^{-5}$\\
O$^+$&2470&25.9&5.05$\times10^{-5}$\\
  &3726&20.8&3.19$\times10^{-5}$\\
  &3729&7.51&3.17$\times10^{-5}$\\
  &7320&17.9&4.72$\times10^{-5}$\\
  &7330&14.7&4.82$\times10^{-5}$\\
&adopted&&4.33$\times10^{-5}$\\
O$^{2+}$&1663&34.5&2.56$\times10^{-4}$\\
  &4363&25.4&3.07$\times10^{-4}$\\
  &4931&0.254&3.33$\times10^{-4}$\\
  &4959&564.0&3.03$\times10^{-4}$\\
  &5007&1657.92&3.08$\times10^{-4}$\\
  &52$\mu$m&6.86& 1.63$\times10^{-4}$\\
  &88$\mu$m&0.79& 1.83$\times10^{-4}$\\
&adopted&&3.06$\times10^{-4}$\\
O$^{3+}$&1400&49.5$^{\mathrm{b}}$ &1.39$\times10^{-4}$\\
        &26$\mu$m&40.2&6.63$\times10^{-5}$\\
&adopted&&6.63$\times10^{-5}$\\
Ne$^+$&12.8$\mu$m&5.33&7.16$\times10^{-6}$\\
Ne$^{2+}$&3869&125.8&5.37$\times10^{-5}$\\
         &36$\mu$m& 2.82&4.25$\times10^{-5}$\\
&adopted&&5.35$\times10^{-5}$\\
Ne$^{3+}$&1602&13.8&3.30$\times10^{-5}$\\
   &2422&120.4&4.32$\times10^{-5}$\\
   &2425&41.3&4.15$\times10^{-5}$\\
   &4714&0.903&2.82$\times10^{-5}$\\
   &4715&0.257 &2.87$\times10^{-5}$\\
   &4724&0.807&2.31$\times10^{-5}$\\
   &4726&0.677&2.22$\times10^{-5}$\\
&   adopted&&4.19$\times10^{-5}$\\
Ne$^{4+}$&3346&38.0&2.02$\times10^{-5}$\\
   &14.3$\mu$m&72.2&1.23$\times10^{-5}$\\
&adopted&&1.50$\times10^{-5}$\\
S$^+$&4068&7.61&5.06$\times10^{-7}$\\
   &4076&2.44&4.85$\times10^{-7}$\\
   &6716&1.72&6.56$\times10^{-7}$\\
 &6731&3.88&6.78$\times10^{-7}$\\
&adopted&&5.62$\times10^{-7}$\\
S$^{2+}$&6312&4.23&3.46$\times10^{-6}$\\
   &8830&0.011&3.51$\times10^{-6}$\\
   &33.5$\mu$m&0.494&2.13$\times10^{-6}$\\
   &18.7$\mu$m&4.69&2.18$\times10^{-6}$\\
&adopted&&2.75$\times10^{-6}$\\
S$^{3+}$&10.5$\mu$m&22.7&1.56$\times10^{-6}$\\
Cl$^{+}$&8578&0.324&2.08$\times10^{-8}$\\
    &9124&0.14&3.40$\times10^{-8}$\\
&adopted&&2.48$\times10^{-8}$\\
 Cl$^{2+}$&3353&0.099&4.55$\times10^{-8}$\\
          &5518&0.189&5.40$\times10^{-8}$\\
  &8434&0.074&7.70$\times10^{-8}$\\
           &8481&0.073&7.81$\times10^{-8}$\\
  &adopted&&5.96$\times10^{-8}$\\
  Cl$^{3+}$&5324&0.06&6.07$\times10^{-8}$\\
          &7531&0.457&5.74$\times10^{-8}$\\
          &8045&1.05&5.69$\times10^{-8}$\\
          &11.8$\mu$m&0.46&1.01$\times10^{-7}$\\
&  adopted&&6.71$\times10^{-8}$\\
 Cl$^{4+}$&6.7$\mu$m&0.365&1.08$\times10^{-8}$\\
 Ar$^+$   &7.0$\mu$m&1.726&1.43$\times10^{-7}$\\
 Ar$^{2+}$&7136&21.4&1.09$\times10^{-6}$\\
          &7751&4.65&9.79$\times10^{-7}$\\
          &9.0$\mu$m&6.26&6.82$\times10^{-7}$\\
          &21.9$\mu$m&0.339&9.20$\times10^{-7}$\\
 &adopted&&9.94$\times10^{-7}$\\
 Ar$^{3+}$&4712&2.21&5.84$\times10^{-7}$\\
          &4741&8.11&6.45$\times10^{-7}$\\
 &adopted&&6.32$\times10^{-7}$\\
 Ar$^{4+}$&4626&0.07&1.63$\times10^{-7}$\\
          &6435&1.31&1.99$\times10^{-7}$\\
          &7.9$\mu$m&2.50&1.31$\times10^{-7}$\\
 &adopted&&1.54$\times10^{-7}$ \\
 Ar$^{5+}$&4.5$\mu$m&6.88&8.58$\times10^{-8}$\\
 Mg$^{3+}$&4.4$\mu$m&8.67&3.65$\times10^{-6}$\\
 Mg$^{4+}$&2783&9.53&1.39$\times10^{-5}$\\
          &2929&3.39&1.27$\times10^{-5}$\\
          & 5.6$\mu$m&22.1&9.88$\times10^{-6}$\\
 &adopted&&1.18$\times10^{-5}$\\
 Fe$^{2+}$&4702&0.067&3.09$\times10^{-8}$\\
          &4734&0.032&2.97$\times10^{-8}$\\
          &4755&0.032&3.88$\times10^{-8}$\\
          &4770&0.023&3.07$\times10^{-8}$\\
          &4778&0.018&3.46$\times10^{-8}$\\
      &4881&0.069$^{\mathrm{c}}$&1.77$\times10^{-8}$\\
          &5270&0.094&3.70$\times10^{-8}$\\
 &adopted&&3.40$\times10^{-8}$\\
 Fe$^{3+}$&6740&0.015&7.08$\times10^{-8}$\\
          &7183&0.016&3.93$\times10^{-7}$\\
          &7189&0.029&3.73$\times10^{-7}$\\
 &adopted&&3.02$\times10^{-7}$\\
 Fe$^{5+}$&5234&0.020&1.19$\times10^{-9}$\\
          &5278&0.035&1.24$\times10^{-9}$\\
          &5335&0.102&1.97$\times10^{-9}$\\
          &5425&0.117&1.56$\times10^{-9}$\\
          &5484&0.070&2.44$\times10^{-9}$\\
          &5631&0.083&1.35$\times10^{-9}$\\
  &adopted&&1.73$\times10^{-9}$\\
 Fe$^{6+}$&4894&0.062&1.05$\times10^{-9}$\\
          &4989&0.089&1.68$\times10^{-9}$\\
          &5159&0.145&1.75$\times10^{-9}$\\
          &adopted&&1.56$\times10^{-9}$\\
\end{longtable}
\begin{list}{}{}
\item[$^{\mathrm{a}}$] Corrected for a $4\%$ contribution from the [\ion{O}{iii}]
  $\lambda$2322 line, assuming [\ion{O}{iii}] $I(\lambda2322)/(\lambda1663)=0.14$;
\item[$^{\mathrm{b}}$] Blended with the \ion{Si}{iv} $\lambda$1400 line;
\item[$^{\mathrm{c}}$] Probably underestimated because of absorption by the
    DIB at 4882.56\,{\AA}.
\end{list}

\begin{longtable}{llcl}
\caption{Ionic abundances from ORLs. \label{orl_ab}}\\
\hline\hline
{Ion} & {$\lambda$} & {$I(\lambda)$}&
{N(X$^{i+}$)/N(H$^+$)} \\
 & {({\AA})} &
{[$I({\rm H}\beta)=100$]} & \\
\hline
\endfirsthead
\caption{continued.}\\
\hline\hline
{Ion} & {$\lambda$} & {$I(\lambda)$}&
{N(X$^{i+}$)/N(H$^+$)} \\
 & {({\AA})} &
{[$I({\rm H}\beta)=100$]} & \\
\hline
\endhead
\hline
\endfoot
He$^{+}$  & 4471 & 3.222& 5.63$\times10^{-2}$\\
          & 6678 & 2.538& 5.82$\times10^{-2}$\\
          & 5876 & 10.87& 5.80$\times10^{-2}$\\
          &adopted &      & 5.80$\times10^{-2}$\\
He$^{2+}$ & 4686 & 47.75& 4.04$\times10^{-2}$\\
          & 4860 & 2.833& 4.39$\times10^{-2}$\\
&adopted         &      & 4.10$\times10^{-2}$\\
C$^{2+}$  & 3918 & 0.024& 1.04$\times10^{-3}$\\
          & 4267 & 0.575& 6.40$\times10^{-4}$\\
          & 4802 & 0.032& 1.01$\times10^{-3}$\\
          & 5342 & 0.034& 7.15$\times10^{-4}$\\
          & 6462 & 0.060& 6.64$\times10^{-4}$\\
          & 6578 & 0.231& 3.99$\times10^{-4}$\\
          & 7231 & 0.108& 2.56$\times10^{-4}$\\
&adopted         &      & 5.54$\times10^{-4}$\\
C$^{3+}$  & 4070 & 0.425& 2.70$\times10^{-4}$\\
          & 4187 & 0.197& 2.70$\times10^{-4}$\\
          & 8196 & 0.289& 6.20$\times10^{-4}$\\
&adopted         &      & 3.87$\times10^{-4}$\\
C$^{4+}$  & 4658 & 0.579$^{\mathrm{a}}$& 2.81$\times10^{-4}$\\
N$^{2+}$  & 3995 & 0.009& 8.49$\times10^{-5}$\\
          & 4041 & 0.012& 8.42$\times10^{-5}$\\
          & 4043 & 0.008& 7.41$\times10^{-5}$\\
          & 4678 & 0.007& 1.18$\times10^{-4}$\\
          & 4179 & 0.005& 7.62$\times10^{-5}$\\
          & 4237 & 0.005& 5.07$\times10^{-5}$\\
          & 4431 & 0.006& 9.75$\times10^{-5}$\\
          & 4631 & 0.021& 8.86$\times10^{-5}$\\
          & 5667 & 0.014& 6.03$\times10^{-5}$\\
          & 5680 & 0.033& 7.60$\times10^{-5}$\\
          & 5686 & 0.019& 2.47$\times10^{-4}$\\
          & 5927 & 0.002& 6.83$\times10^{-5}$\\
          & 5940 & 0.004& 1.37$\times10^{-4}$\\
          & 6482 & 0.004& 9.90$\times10^{-5}$\\
&adopted         &      & 1.03$\times10^{-4}$\\
N$^{3+}$  & 4379 & 0.059$^{\mathrm{b}}$& 2.56$\times10^{-5}$\\
N$^{4+}$  & 3483 & 0.031& 1.22$\times10^{-5}$\\
          & 4707 & 0.015& 2.34$\times10^{-5}$\\
          & 7582 & 0.011& 2.29$\times10^{-5}$\\
          & 7703 & 0.021& 1.66$\times10^{-5}$\\
&adopted         &      & 1.68$\times10^{-5}$\\
N$^{5+}$  & 4945 & 0.043& 7.02$\times10^{-6}$\\
O$^{2+}$  & 4072 & 0.134& 5.55$\times10^{-4}$\\
          & 4111 & 0.012& 5.04$\times10^{-4}$\\
          & 4119 & 0.035& 3.99$\times10^{-4}$\\
          & 4133 & 0.013& 2.40$\times10^{-4}$\\
          & 4153 & 0.021& 2.72$\times10^{-4}$\\
          & 4317 & 0.022& 2.80$\times10^{-4}$\\
          & 4320 & 0.026& 3.06$\times10^{-4}$\\
          & 4346 & 0.017& 2.17$\times10^{-4}$\\
          & 4349 & 0.041& 2.08$\times10^{-4}$\\
          & 4367 & 0.035& 4.41$\times10^{-4}$\\
          & 4639 & 0.070& 6.38$\times10^{-4}$\\
          & 4649 & 0.165& 3.13$\times10^{-4}$\\
          & 4662 & 0.036& 2.56$\times10^{-4}$\\
          & 4676 & 0.037& 3.14$\times10^{-4}$\\
&adopted         &      & 3.95$\times10^{-4}$\\
O$^{3+}$  & 4435 & 0.019& 4.04$\times10^{-5}$\\
          & 5592$^{\mathrm{c}}$ & 0.105& 6.20$\times10^{-5}$\\
&adopted         &      & 4.04$\times10^{-5}$\\
O$^{4+}$  & 3410 & 0.103& 1.43$\times10^{-5}$\\
          & 4632 & 0.188& 1.19$\times10^{-5}$\\
&adopted         &      & 1.27$\times10^{-5}$\\
O$^{5+}$  & 7611 & 0.014& 7.46$\times10^{-6}$\\
Ne$^{2+}$ & 3327 & 0.016& 8.86$\times10^{-5}$\\
          & 3335 & 0.042& 6.24$\times10^{-5}$\\
          & 3355 & 0.027& 7.68$\times10^{-5}$\\
          & 3694 & 0.023& 6.78$\times10^{-5}$\\
          & 3777 & 0.011& 8.35$\times10^{-5}$\\
          & 4392 & 0.011& 8.43$\times10^{-5}$\\
          & 4409 & 0.017& 1.96$\times10^{-4}$\\
&adopted         &      & 8.63$\times10^{-5}$\\
Mg$^{2+}$ & 4481 & 0.023& 2.43$\times10^{-5}$\\
\end{longtable}
\begin{list}{}{}
\item[$^{\mathrm{a}}$]Corrected for a $23\%$ contribution from the [\ion{Fe}{iii}]
  $\lambda$4658 line, assuming [\ion{Fe}{iii}] $I(\lambda4658)/I(\lambda4734)=5.33$;
\item[$^{\mathrm{b}}$]Corrected for a total of $11\%$ contribution from the
\ion{C}{iii} and \ion{Ne}{ii} $\lambda$4380 lines, assuming
\ion{C}{iii} $I(\lambda4380)/(\lambda4383)=3.5$ and
\ion{Ne}{ii} $I(\lambda4380)/(\lambda4392)=0.61$;
\item[$^{\mathrm{c}}$] Also excited partly by charge exchange with H$^0$ \citep{liu1993}.
\end{list}

\Online

\setcounter{figure}{0}
\begin{figure*}
\centering
\epsfig{file=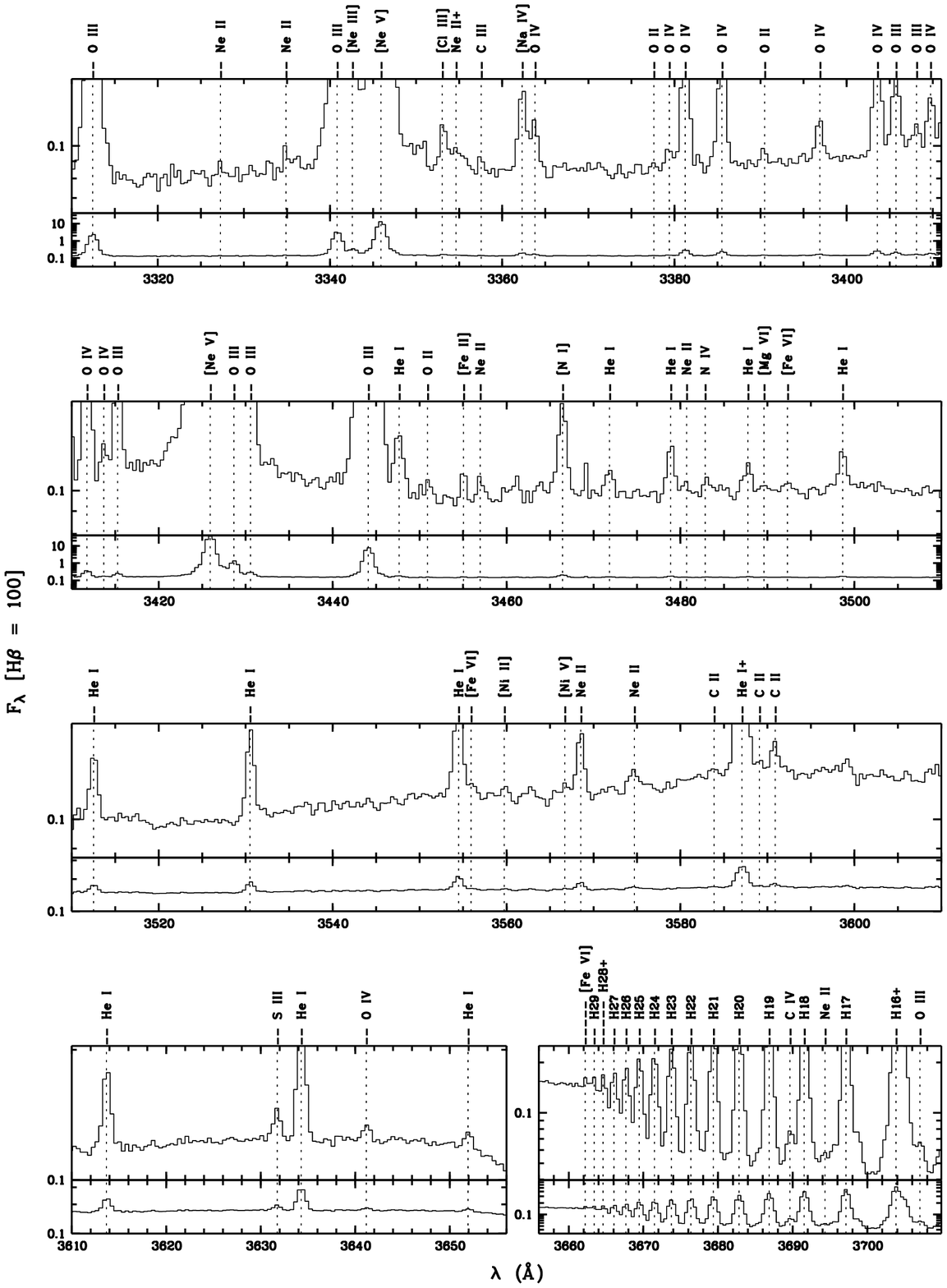,
height=22cm, bbllx=44, bblly=54, bburx=550, bbury=758, clip=, angle=0}
\caption{Optical spectrum of NGC 7027. A `+' attached to a line
identification means that the line is blended with other emission features.}
\label{all}
\end{figure*}

\begin{figure*}
\centering
\epsfig{file=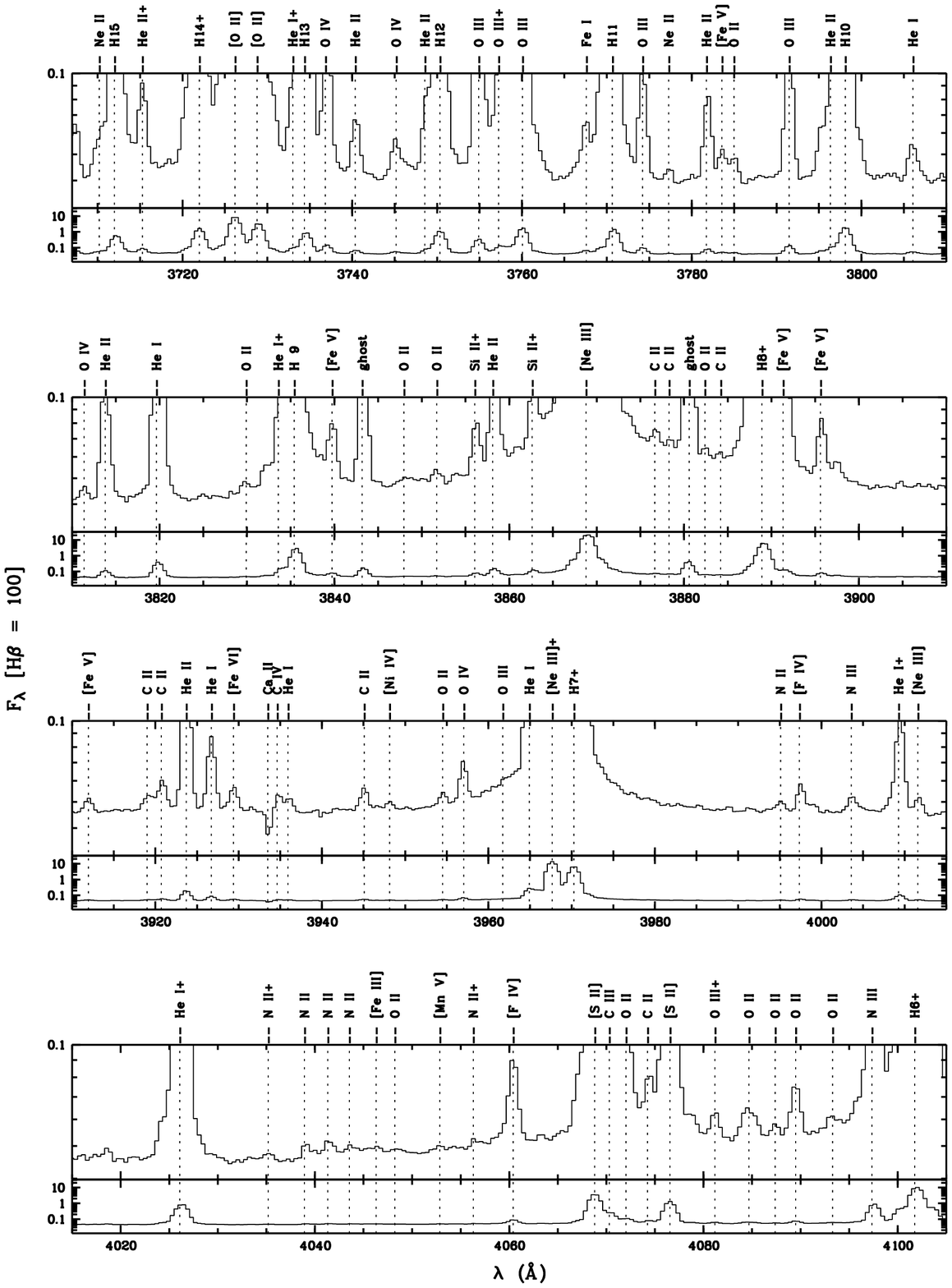,
height=22cm, bbllx=44, bblly=54, bburx=550, bbury=758, clip=, angle=0}
\end{figure*}

\begin{figure*}
\centering
\epsfig{file=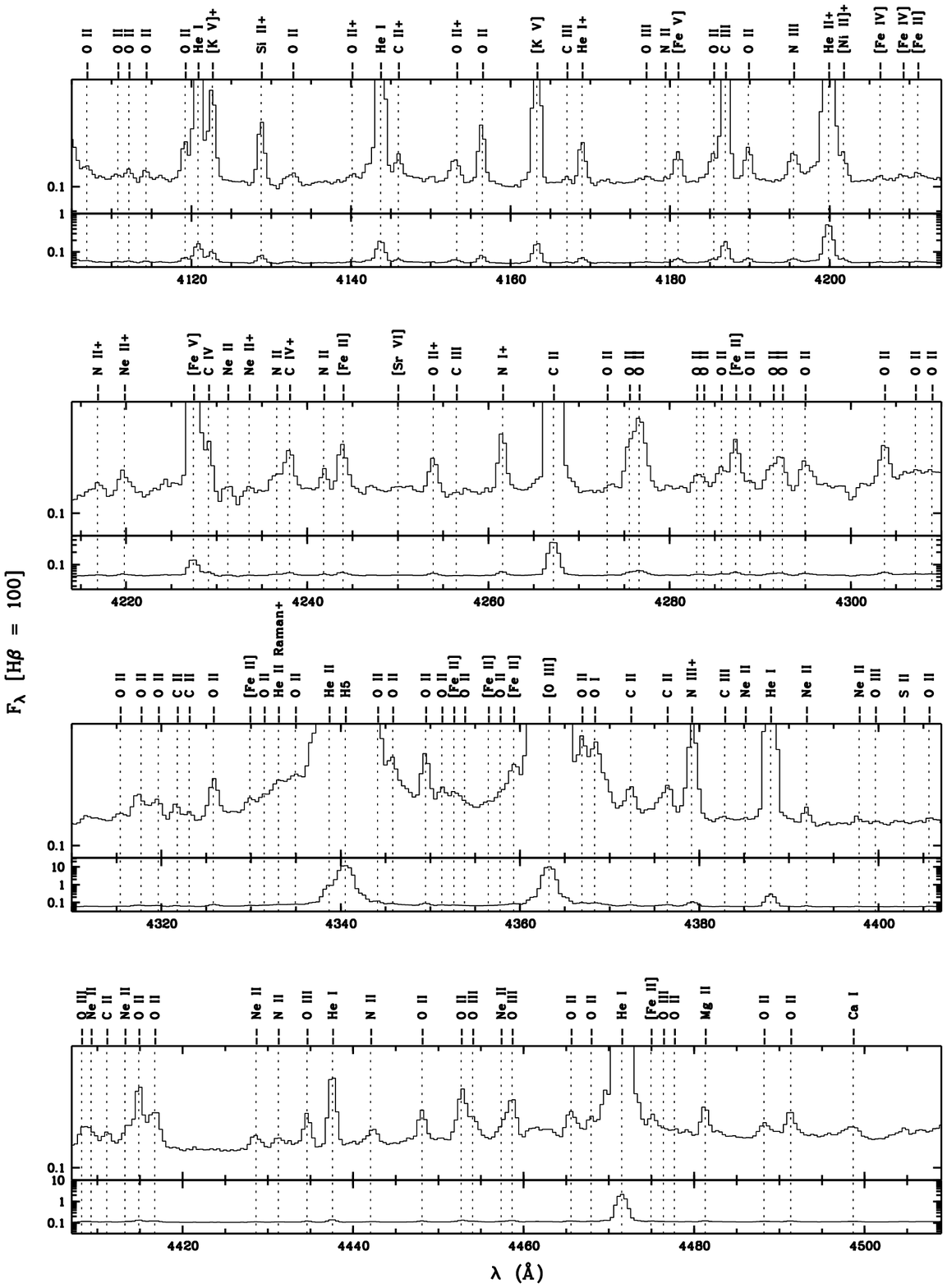,
height=22cm, bbllx=44, bblly=54, bburx=550, bbury=758, clip=, angle=0}
\end{figure*}

\begin{figure*}
\centering
\epsfig{file=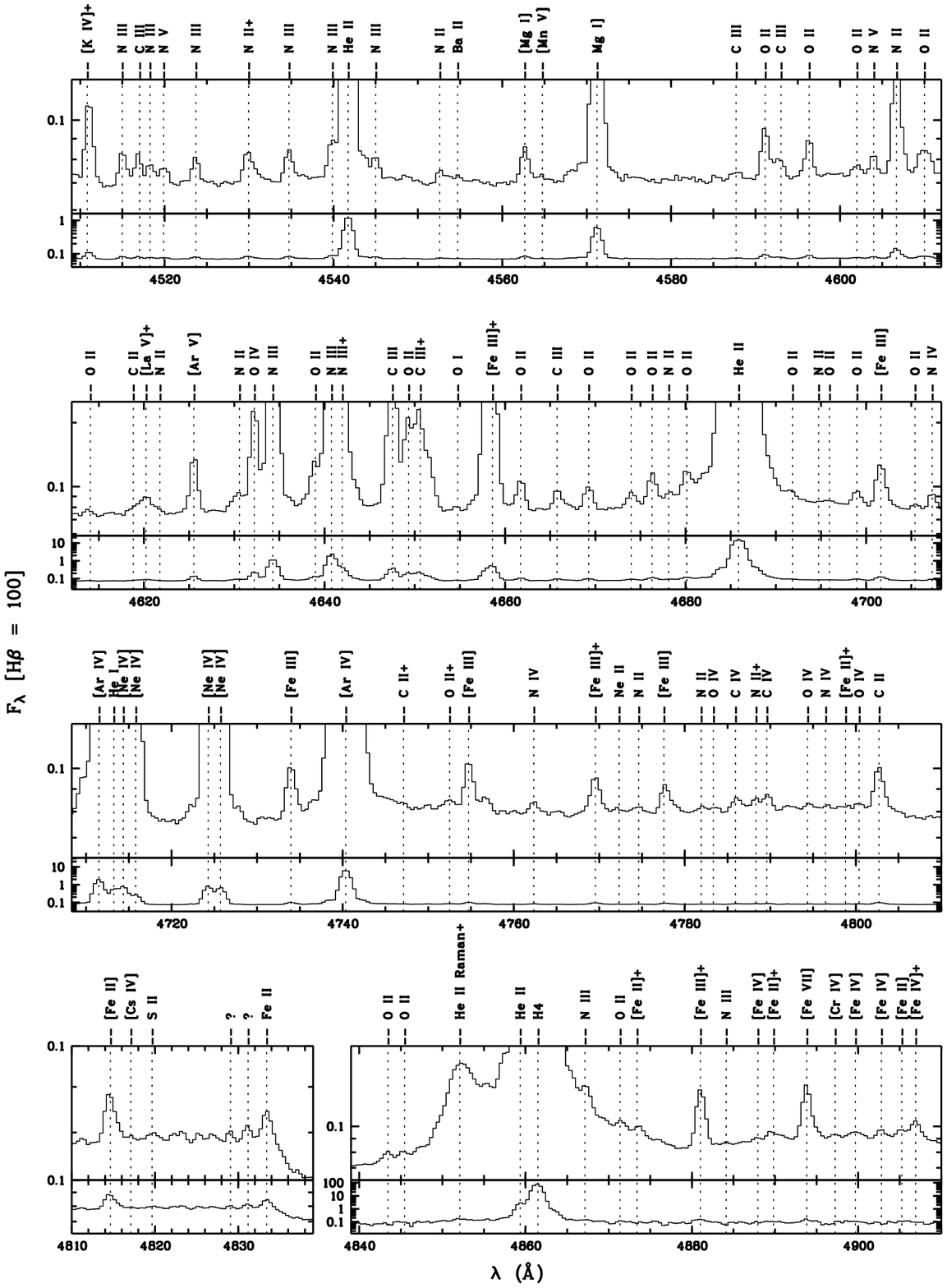,
height=22cm, bbllx=44, bblly=54, bburx=550, bbury=758, clip=, angle=0}
\end{figure*}

\begin{figure*}
\centering
\epsfig{file=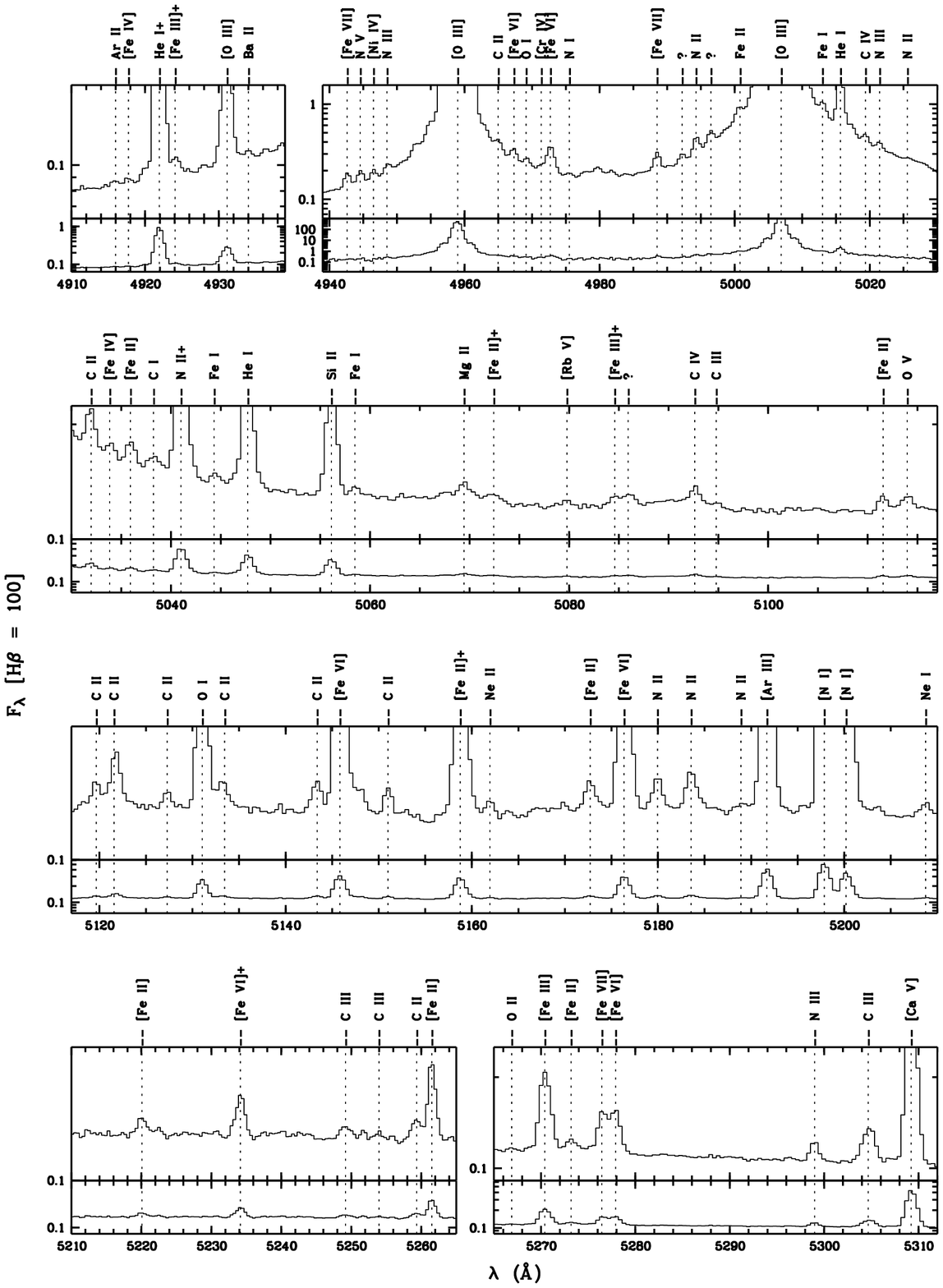,
height=22cm, bbllx=44, bblly=54, bburx=550, bbury=758, clip=, angle=0}
\end{figure*}

\begin{figure*}
\centering
\epsfig{file=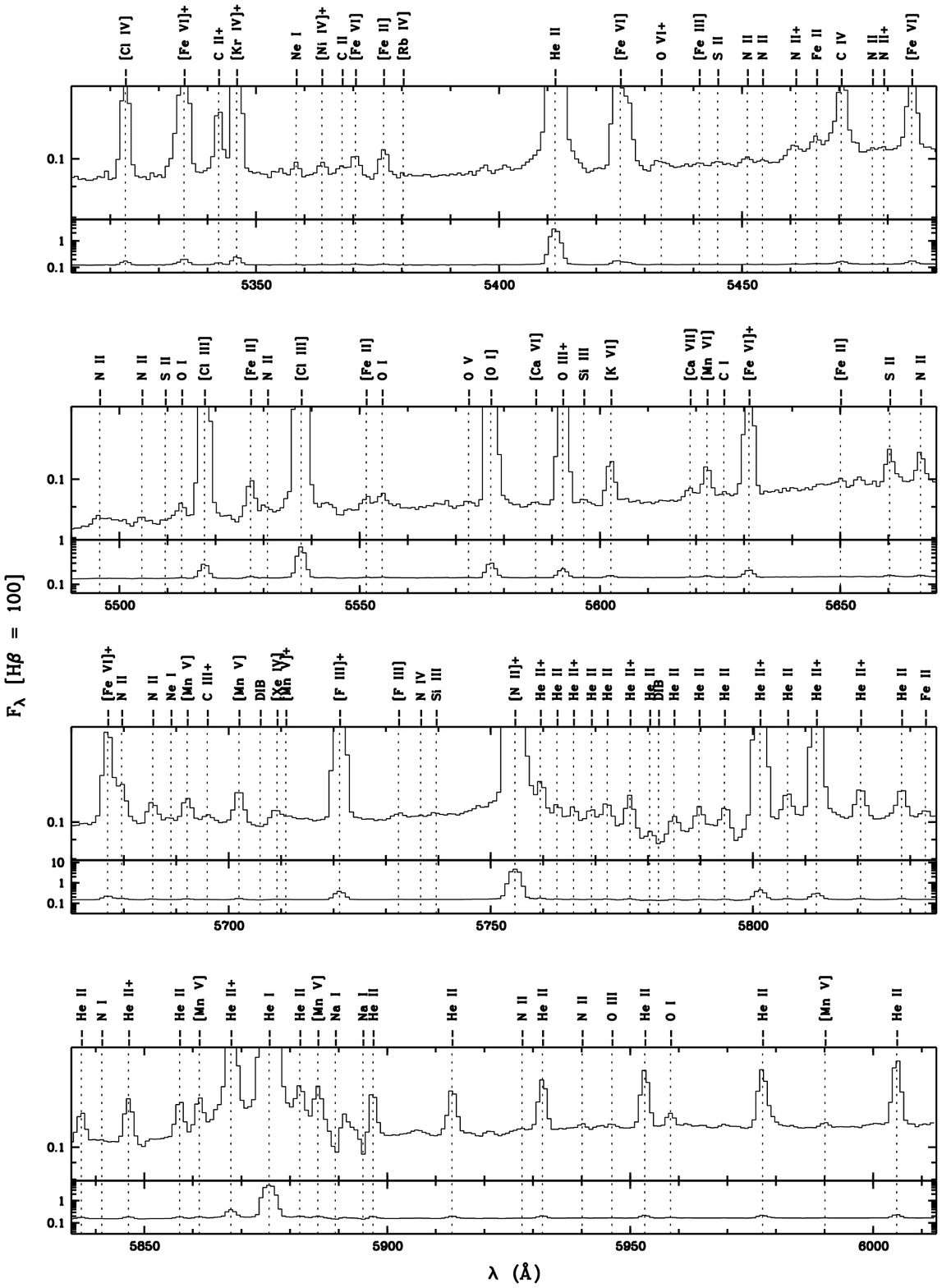,
height=22cm, bbllx=44, bblly=54, bburx=550, bbury=758, clip=, angle=0}
\end{figure*}

\begin{figure*}
\centering
\epsfig{file=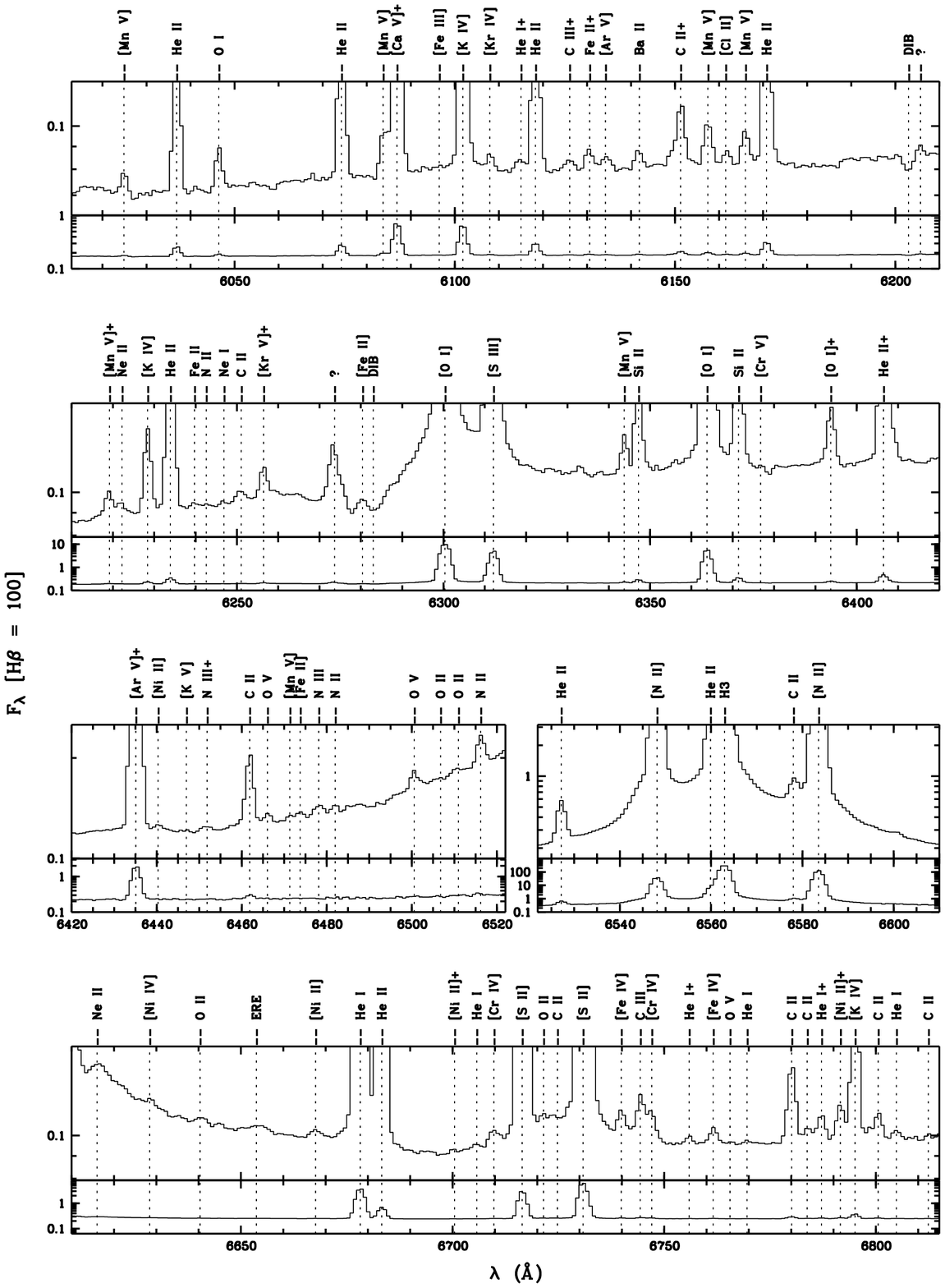,
height=22cm, bbllx=44, bblly=54, bburx=550, bbury=758, clip=, angle=0}
\end{figure*}

\begin{figure*}
\centering
\epsfig{file=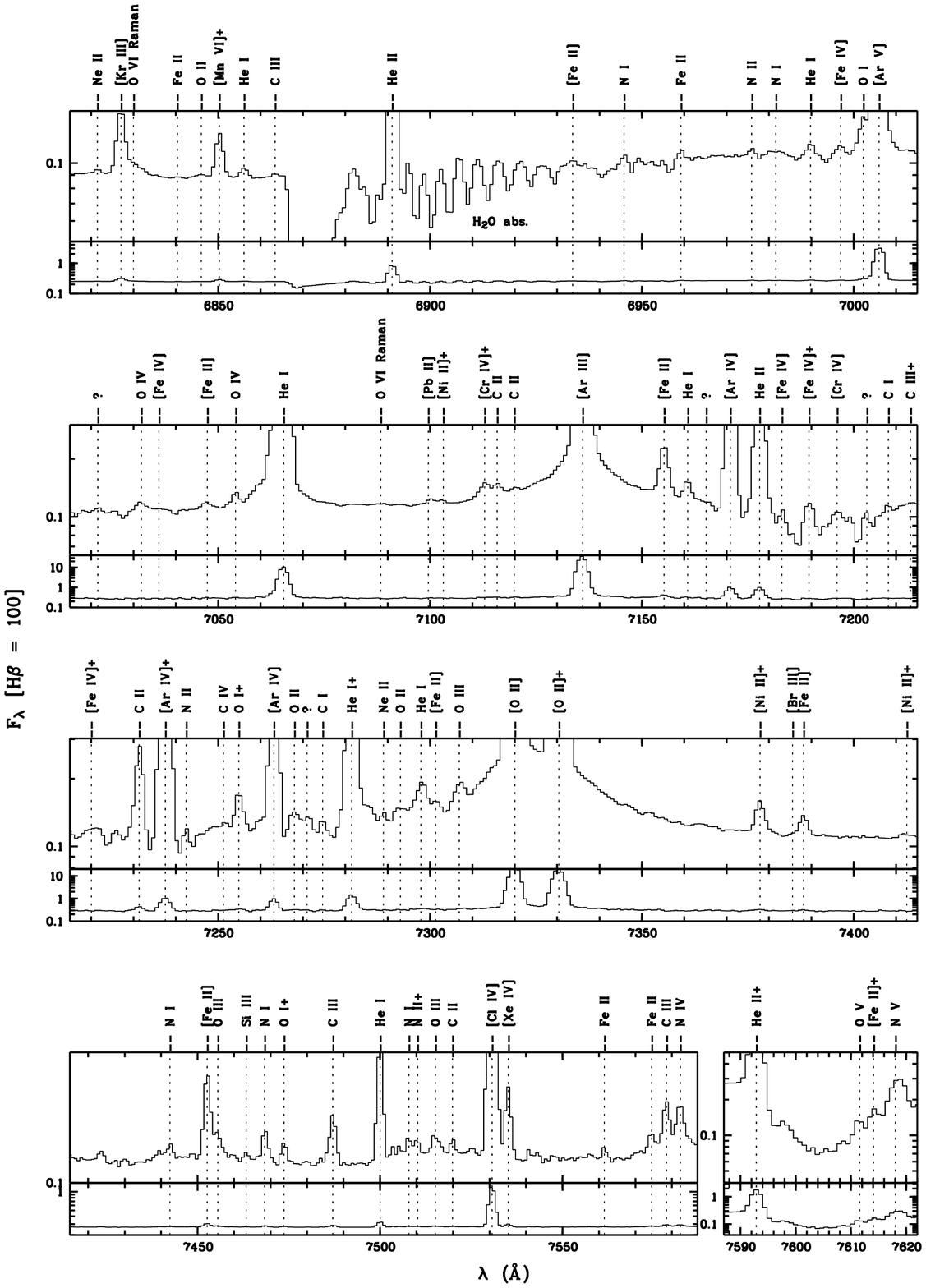,
height=22cm, bbllx=44, bblly=54, bburx=550, bbury=758, clip=, angle=0}
\end{figure*}

\begin{figure*}
\centering
\epsfig{file=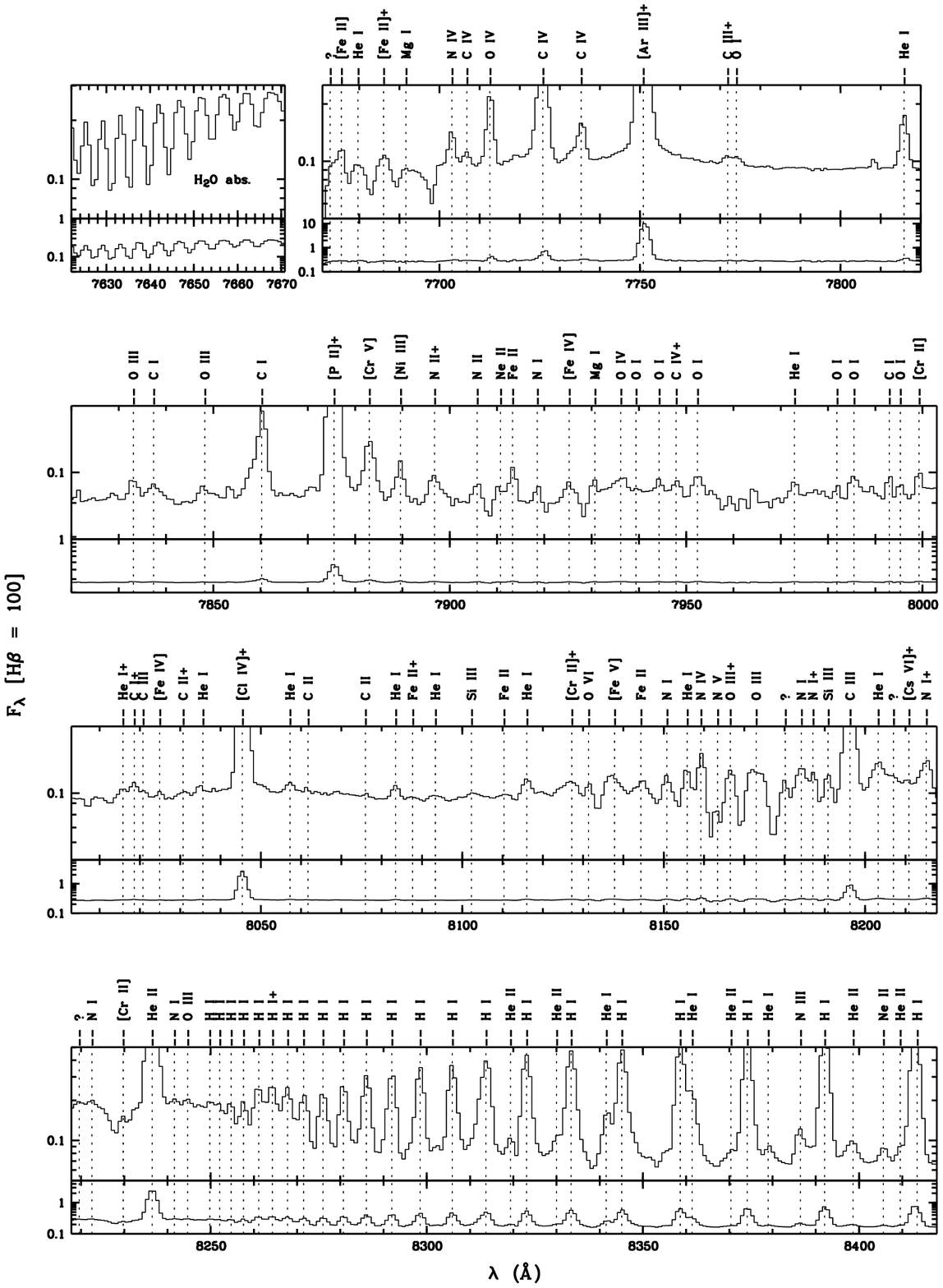,
height=22cm, bbllx=44, bblly=54, bburx=550, bbury=758, clip=, angle=0}
\end{figure*}

\begin{figure*}
\centering
\epsfig{file=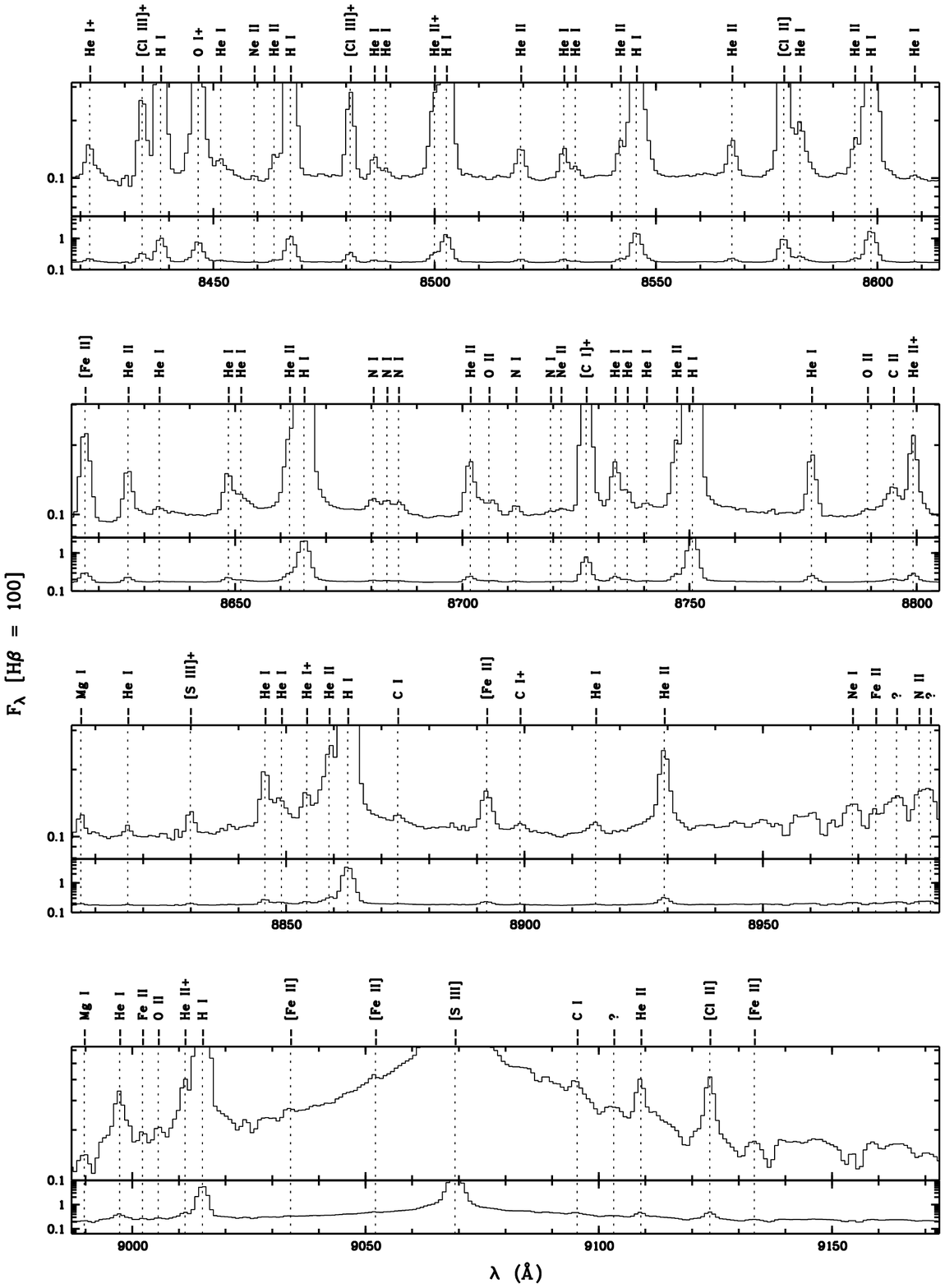,
height=22cm, bbllx=44, bblly=54, bburx=550, bbury=758, clip=, angle=0}
\end{figure*}

\clearpage

\setcounter{table}{0}
\begin{table}
\centering
\caption{Journal of WHT observations.\label{jan}}
\begin{tabular}{cccc}
\hline\hline
Date & $\lambda$-range   & $\Delta\lambda$& Exp. Time\\
(UT)& ({\AA}) & {({\AA})} & {(sec)}\\
\hline
     1996/08/23&3620--4400&2.5&$4\times14$, 1200\\
     1996/08/23&4200--4980&2.5&10, $2\times7$, $2\times20$,$2\times600$\\
     1996/08/23&5200--6665&4.5&10, $2\times7$, $2\times20$,$2\times600$\\
     1996/08/23&6460--7930&4.5&$4\times14$, 1200\\
     1997/08/10&3310--3710&1.5&$2\times1200$\\
     1997/08/10&3710--4105&1.5&20, $2\times1200$\\
     1997/08/10&4110--4505&1.5&20, $3\times1200$\\
    1997/08/10&4510--4905&1.5&20, 25, 600, $4\times1200$\\
    1997/08/10&4910--5310&1.5&10, $2\times20$, 300, $4\times600$\\
    1997/08/10&5220--6220&2.5&20, $2\times1200$\\
    1997/08/10&6020--6800&2.5&10, $2\times20$, 300, $4\times600$\\
    1997/08/10&6770--7650&2.5&20, $3\times1200$\\
    1997/08/10&7620--8410&2.5&20, 25, 600, $4\times1200$\\
    1997/08/10&8370--9160&2.5&$2\times1200$\\
\hline\end{tabular}
\end{table}

\setcounter{table}{2}
\begin{table}
\begin{center}
\caption{Journal of {\it IUE}\ observations.\label{iuejan}}
\begin{tabular}{ccc}
\hline\hline
{Date}           & {Image No.}      &
{Exp. Time}   \\
{(UT)} &  &{(min)} \\
\hline
1982/06/17& L SWP 17242 L& 180\\
1983/05/02& L SWP 19878 L& 162\\
1978/07/04& L SWP 01914 L&  36\\
1978/09/01& L SWP 02430 L&  30\\
1978/01/08& L SWP 01747 L&  10\\
\noalign{\vskip5pt}
1978/03/29& H SWP 19578 L& 420\\
1979/03/21& H SWP 04716 L& 287\\
1979/09/17& H SWP 06541 L&  60\\
1982/06/17& H SWP 17240 L&  45\\
1979/03/25& H SWP 04748 L&  35\\
\noalign{\vskip5pt}
1983/05/02& L LWR 15861 L& 180\\
1978/02/24& L LWR 01024 L& 120\\
1979/09/17& L LWR 05614 L&  60\\
1982/06/17& L LWR 13506 L&  60\\
1978/11/01& L LWR 02785 L&  30\\
1978/06/08& L LWR 01639 L&  25\\
1978/09/01& L LWR 02230 L&  10\\
\noalign{\vskip5pt}
1983/01/25& H LWR 15105 L& 300\\
1979/09/17& H LWR 05615 L&  35\\
\hline\end{tabular}
\end{center}
\end{table}

\appendix

\section{Emission from heavy elements ($Z>30$)}

\citet{pequignotba1994} identify emission lines from a number of heavy elements
with $Z>30$ in \object{NGC\,7027} for the first time. As expected, many of
these lines have extremely low intensities [$\sim 10^{-5}I({\rm H}\beta)$].
Table~\ref{rs} lists emission lines from heavy elements of $Z>30$ identified by
\citet{pequignotba1994} (PB) that have also been detected in our optical spectrum.
The measured fluxes are in general consistent with those reported by
\citet{pequignotba1994}.  These emission features should provide vital
information about r- and s-processes in late evolution stages of AGB stars.

\begin{table}
\begin{center}
\caption{Emission lines from elements of $Z>30$. \label{rs}} 
\begin{tabular}{llcc}
\hline\hline
{Ion} & {$\lambda$({\AA})} &     \multicolumn{2}{c}{$10^5I/I({\rm H}\beta)$} \\
\cline{3-4}\\
    &          &     {This}  &  {PB}\\
\hline
{[Se~{\sc ii}]}   &    7592.00    &             blend$^a$    &     blend              \\
{[Se~{\sc iii}]}  &    8854.20    &             $<9$     &     $15.0\pm2.3$   \\
{[Br~{\sc iii}]}  &    6131.00    &             $<11$    &     $7.6\pm1.1$    \\
{[Br~{\sc iii}] } &    7385.1     &              2       &     $<2.5$         \\
{[Kr~{\sc iii}]} &    6826.90    &             44$^b$   &    $41\pm2$        \\
{[Kr~{\sc iv}] } &    5346.10    &           148$^c$    &     $187\pm6$      \\
{[Kr~{\sc iv}] } &    5868.00    &           231$^d$    &     $255\pm15$     \\
{[Kr~{\sc iv}] }  &    6107.80    &              5       &     $6.3\pm1.6$    \\
{[Kr~{\sc v}]  }  &    6256.50    &            18$^e$    &     $15.1\pm1.5$   \\
{[Rb~{\sc iv}] }  &    5759.40    &             23$^f$   &     $17.4\pm2.6$   \\
{[Rb~{\sc v}]  }  &    5080.20    &              7       &     $<4$           \\ 
{[Sr~{\sc vi}] }  &    4249.30    &              1       &     $7\pm5$        \\
{[Sr~{\sc vi}] }  &    5434.40    &            $<6$      &     $8\pm4$        \\
{[Sr~{\sc v}]  }  &    4922.20    &              blend       &      blend             \\
{[Zr~{\sc vii}]}  &    7379.70    &            $<35$     &       blend            \\
{[Te~{\sc ii}] }  &    8049.60    &              blend       &     $<1.5$        \\
{[I~{\sc ii}]  }  &    7282.90    &              blend       &      blend           \\
{[Xe III]      } &    5846.70    &             12$^g$   &     $9.3\pm2$    \\
{[Xe~{\sc iv}] }  &    5709.20    &              9       &     $10.6\pm2$    \\
{[Xe~{\sc iv}] }  &    7535.40    &             14       &     $15.8\pm0.6$  \\
{[Xe~{\sc vi}] }  &    6408.89    &              blend       &     $13.2\pm1.6$  \\
{[Cs~{\sc ii}] }  &    7219.70    &            $<19$     &     $<10$         \\
{[Cs~{\sc vi}] }  &    8210.60    &            $<14$     &     $3\pm2$       \\
Ba~{\sc ii}     &    4554.00    &              2       &      7.8:         \\
Ba~{\sc ii}     &    4934.10    &              4       &      $<5.6$       \\
Ba~{\sc ii}     &    6141.70    &              7       &      7.5          \\
{[Ba~{\sc iv}]  } &    5696.60    &             8$^h$    &     $7.3\pm2.2$   \\ 
{[Ba~{\sc viii}]}&    4233.60    &            $<3$      &     $5\pm4$       \\
{[Pb~{\sc ii}]  }&    7099.80    &              10      &     $3.5\pm1.0$   \\
{[La~{\sc v}]   }&    4621.00    &            $<26$     &       blend          \\
\hline\end{tabular}

\begin{list}{}{}
\item[$^{\mathrm{a}}$]Blended with a strong line;
\item[$^{\mathrm{b}}$]Ignore contribution from the C~{\sc I} $\lambda$6828.11 line;
\item[$^{\mathrm{c}}$]Ignore contribution from the C~{\sc III}$\lambda$5345.80 line;
\item[$^{\mathrm{d}}$]Corrected for a $9\%$ contribution from the He~{\sc ii} $\lambda$5869.00 line,
assuming He~{\sc II} $I(\lambda5869.00)/I(\lambda5896.79)=0.58$;
\item[$^{\mathrm{e}}$]Corrected for a $28\%$ contribution from the C~{\sc ii} $\lambda$6256.54 line,
assuming C~{\sc ii} $I(\lambda6256.54)/I(\lambda6250.74)=0.56$;
\item[$^{\mathrm{f}}$] Corrected for a $18\%$ contribution from the He~{\sc ii}
$\lambda$5759.44 line, assuming He~{\sc ii} $I(\lambda5759.44)/I(\lambda5762.63)=0.71$;
\item[$^{\mathrm{g}}$]Corrected for a $69\%$ contribution from the He~{\sc ii} $\lambda$5847.10 line,
assuming He~{\sc II} $I(\lambda 5847.10)/I(\lambda5857.27)=0.93$;
\item[$^{\mathrm{h}}$]Neglect contribution from the C~{\sc iii} $\lambda$5695.9 line.
\end{list}
\end{center}
\end{table}

\section{Atomic data references}

Table~\ref{refcel} and Table~\ref{reforl} list references of atomic data for
CEL and ORL analyses, respectively.

\begin{table*}
\centering
\caption{Atomic data references for CEL analysis.\label{refcel}}
\begin{tabular}{lll}
\hline\hline
{Ion} & {Transition probabilities} &{Collision strengths}\\
\hline
C~{\sc ii}  & \citealt{Nussbaumer1981}  & \citealt{Blum1992} \\
C~{\sc iii} & \citealt{Keenan1992}      & \citealt{Keenan1992} \\
            & \citealt{Fleming1996}     &                     \\
C~{\sc iv}  & \citealt{wiese}           & \citealt{gau} \\
N~{\sc ii}  & \citealt{Nussbaumer1979}  & \citealt{stafford1994b} \\
N~{\sc iii} & \citealt{Fang1993}        & \citealt{Blum1992} \\
            &                           & \citealt{stafford1994a} \\
\hline\end{tabular}
\end{table*}

\begin{table*}
\noindent \parbox{14cm}{
\centering
\caption{Atomic data references for ORL analysis.\label{reforl}}
\begin{tabular}{lll}
\hline\hline
{Ion} & {Effective recomb. coeffs.} &{Comment} \\
\hline
H\,{\sc i}  & \citealt{storey1995}      & Case B    \\
He\,{\sc i} & \citealt{benjamin}    & Case B; singlets \\
            & \citealt{brocklehurst1972}& Case A; triplets \\
He\,{\sc ii}& \citealt{storey1995}      & Case B    \\
C\,{\sc ii} & \citealt{davey2000}       & Case B    \\
C\,{\sc iii}& \citealt{pequignot1991}   & Case A    \\
            & \citealt{nussbaumer1984}  & Dielectronic recom.\\
C\,{\sc iv} & \citealt{pequignot1991}   & Case A    \\
N\,{\sc ii} & \citealt{kisielius2002}   & Case B; triplets \\
            & \citealt{escalante1990}   & Case A; singlets \\
N\,{\sc iii}& \citealt{pequignot1991}   & Case A    \\
            & \citealt{nussbaumer1984}  & Dielectronic recom.\\
N\,{\sc iv} & \citealt{pequignot1991}   & Case A    \\
            & \citealt{nussbaumer1984}  & Dielectronic recom.\\
N\,{\sc v}  & \citealt{pequignot1991}   & Case A    \\
O\,{\sc ii} & \citealt{storey94}        & Case B; quartets \\
            & \citealt{liu1995}         & Case A; doublets \\
O\,{\sc iii}& \citealt{pequignot1991}   & Case A    \\
O\,{\sc iv} & \citealt{pequignot1991}   & Case A    \\
            & \citealt{nussbaumer1984}  & Dielectronic recom.\\
O\,{\sc v}  & \citealt{pequignot1991}   & Case A    \\
Ne\,{\sc ii}& \citealt{kisielius1998}   & Case B; doublets \\
            & Storey (unpublished)      & Case A; quartets \\
Mg\,{\sc ii} & \citealt{davey2000}$^a$ & Case B    \\
\hline\end{tabular}
    \begin{list}{}{}
\item[$^{\mathrm{a}}$] Given the similarity between the atomic structure of \ion{Mg}{ii} and \ion{C}{ii},
we have assumed that the \ion{Mg}{ii} 3d--4f $\lambda$4481 line has an
effective recombination coefficient equal to that of the \ion{C}{ii} 3d--4f
$\lambda$4267 line.
\end{list}}
\end{table*}

\setcounter{table}{1}
\renewcommand{\thetable}{\arabic{table}}

\begin{longtable}{cccccccccc}
\caption{Optical line fluxes on a scale where H$\beta=100$.\label{opline}}\\
\hline\hline
{$\lambda_{\rm obs}$} & {$F(\lambda)$}   
& {$I(\lambda)$} &{Ion} & {$\lambda_{\rm lab}$} & 
{Mult} &{Lower Term}&{Upper Term}& {$g_1^a$}&{$g_2^a$} \\
\hline
\endfirsthead
\caption{continued.}\\
\hline\hline
{$\lambda_{\rm obs}$} & {$F(\lambda)$}
& {$I(\lambda)$} &{Ion} & {$\lambda_{\rm lab}$} &
{Mult} &{Lower Term}&{Upper Term}& {$g_1^a$}&{$g_2^a$} \\
\hline
\endhead
\hline
\endfoot
 3312.43&     2.385&     7.760&  O III   &  3312.30& V3     &   3s  3P* &  3p  3S &  3&  3\\    
 3327.27&    0.005:&     0.017&  Ne II   &  3327.15& V2     &   3s  4P  &  3p  4D*&  4&  4\\    
 3334.88&     0.014&     0.045&  Ne II   &  3334.84& V2     &   3s  4P  &  3p  4D*&  6&  8\\    
 3340.82&     3.345&    10.547&  O III   &  3340.74& V3     &   3s  3P* &  3p  3S &  5&  3\\    
 3342.58&     0.277&     0.869&  [Ne III]&  3342.50&        &   2p4 1D  &  2p4 1S &  5&  1\\    
 3345.91&    12.858&    40.282&  [Ne V]  &  3345.83&        &   2p2 3P  &  2p2 1D &  3&  5\\    
 3353.44&     0.034&     0.105&  [Cl III]&  3353.33&        &   3p3 4S* &  3p3 2P*&  4&  2\\    
 3354.62&     0.028&     0.086&  Ne II   &  3355.05& V2     &   3s  4P  &  3p  4D*&  4&  6\\
\end{longtable}
\begin{list}{}{}
\item[$^{\mathrm{a}}$] For a feature blended with a few lines from the same multiplet,
this is a sum of statistical weights.
\item[$^{\mathrm{b}}$] Saturated even if in short exposures.
\end{list}


\clearpage
\setcounter{table}{3}

\begin{table}
\caption{UV line fluxes.\label{iueline}}
\begin{tabular}{cccc}
\hline\hline
{Wavelength} & {Ion}  & {$F(\lambda)$}$^{\mathrm{a}}$     & {$I(\lambda)$}\\
{(\AA)}  &  &  &
{[$I({\rm H}\beta)=100$]}      \\
\hline
1240 &   N V          &0.15 &   35.94  \\
1335 &   C II         &0.09 &   10.56  \\
1400 &   O IV]+Si IV  &0.60 &   49.54  \\
1485 &   N IV]        &0.65 &   42.28  \\
1548 &   C IV         &12.5 &  695.97  \\
1551 &   C IV         &7.77 &  430.00  \\
1573 &   C III+[Ne V] &0.21 &   11.10  \\
1602 &   [Ne IV]      &0.27 &   13.80  \\
1640 &   He II        &4.40 &  211.23  \\
1663 &   O III]       &0.74 &   34.51  \\
1750 &   N III]       &0.60 &   27.06  \\
1883 &   [Si III]      &0.04 &    2.29  \\
1892 &   Si III]      &0.05 &    3.00  \\
1906 &   [C III]       &4.40 &  280.75  \\
1908 &   C III]       &7.00 &  446.49  \\
2324 &   [O III]+CII] &1.90 &  185.00  \\
2422 &   [Ne IV]      &1.20 &   54.14  \\
2425 &   [Ne IV]      &0.45 &   19.96  \\
2470 &   [O II]       &0.76 &   25.73  \\
2512 &   He II        &0.25 &    6.65  \\
2732 &   He II        &0.56 &    6.68  \\
2783 &   [Mg V]       &0.92 &    9.53  \\
2796 &   Mg II        &1.04 &   10.46  \\
2804 &   Mg II        &1.68 &   16.56  \\
2830 &   He I         &0.16 &    1.48  \\
2929 &   [Mg V]       &0.45 &    3.39  \\
3024 &   O III        &0.87 &    5.53  \\
3046 &   O III        &1.33 &    8.19  \\
3134 &   O III        &9.71 &   52.30  \\
3203 &   He II        &3.05 &   14.95  \\
\hline\end{tabular}
\begin{list}{}{}
\item[$^{\mathrm{a}}$] 10$^{-12}$\,erg\,cm$^{-2}$\,s$^{-1}$.
\end{list}
\end{table}

\end{document}